\documentclass[12pt]{article}
\usepackage{amsmath}
\usepackage{amssymb}
\usepackage{graphicx}
\usepackage{ifthen}
\usepackage{verbatim}

\newlength{\captionwidth}
\newsavebox{\tempbox}
\newcommand{\mycaption}[2]{%
\par\vspace{10pt}\sbox{\tempbox}{Figure #1: #2}%
\ifthenelse{\lengthtest{\wd\tempbox>\captionwidth}}%
{\sbox{\tempbox}{Figure.#1:\ }%
\addtolength{\captionwidth}{-\wd\tempbox}%
\mbox{Figure #1:\ }\parbox[t]{\captionwidth}{\small\textit{#2}}}%
{Figure #1: {\small\textit{#2}}}}%

\numberwithin{equation}{section}

\newtheorem{formula}{Formula}

\allowdisplaybreaks
\begin{document}
\thispagestyle{empty}

\begin{flushright}
July  2008\\
September 2008 (revised)
\end{flushright}
\bigskip
\bigskip

\begin{center}
{\Large Extended $5d$ Seiberg-Witten Theory}\\
{\Large and}\\
{\Large Melting Crystal}
\end{center}
\bigskip
\bigskip

\renewcommand{\thefootnote}{\fnsymbol{footnote}}
\begin{center}
Toshio Nakatsu
\footnote{E-mail: \texttt{nakatsu@phys.sci.osaka-u.ac.jp}}$^1$, 
Yui Noma 
\footnote{E-mail: \texttt{yuhii@het.phys.sci.osaka-u.ac.jp}}$^1$
and 
Kanehisa Takasaki
\footnote{E-mail: \texttt{takasaki@math.h.kyoto-u.ac.jp}}$^2$\\
\bigskip
{\small
\textit{$^1$Department of Physics, Graduate School of Science,
Osaka University,\\
Toyonaka, Osaka 560-0043, Japan}}\\
{\small
\textit{$^2$Graduate School of Human and Environmental Studies, 
Kyoto University,\\ 
Yoshida, Sakyou, Kyoto 606-8501, Japan}}
\end{center}
\bigskip
\bigskip
\renewcommand{\thefootnote}{\arabic{footnote}}

\begin{abstract}
We study an extension of the Seiberg-Witten theory 
of $5d$ $\mathcal{N}=1$ supersymmetric Yang-Mills 
on $\mathbb{R}^4 \times S^1$. 
We investigate correlation functions among loop operators. 
These are the operators analogous to the Wilson loops  
encircling the fifth-dimensional circle 
and give rise to physical observables of 
topologically-twisted $5d$ $\mathcal{N}=1$ 
supersymmetric Yang-Mills 
in the $\Omega$ background. 
The correlation functions are computed 
by using the localization technique. 
Generating function of the correlation functions 
of $U(1)$ theory 
is expressed as a statistical sum over partitions 
and reproduces the partition function of 
the melting crystal model with external potentials. 
The generating function becomes 
a $\tau$ function of $1$-Toda hierarchy, 
where the coupling constants of the loop operators 
are interpreted as time variables of $1$-Toda hierarchy. 
The thermodynamic limit of the partition function 
of this model is studied. 
We solve a Riemann-Hilbert problem 
that determines the limit shape of the main diagonal 
slice of random plane partitions in the presence of 
external potentials, 
and identify a relevant complex curve 
and the associated Seiberg-Witten differential.  
\end{abstract}

\setcounter{footnote}{0}
\newpage

\section{Introduction}

The notion of {\it random partitions} has played 
a central role in recent studies on supersymmetric gauge theories.  
A highlight will be the work of Nekrasov and Okounkov 
\cite{Nekrasov-Okounkov} who used random partitions 
to derive the Seiberg-Witten solution \cite{Seiberg-Witten} 
of $4d$ $\mathcal{N}=2$ supersymmetric gauge theories.  
Random partitions emerge therein through Nekrasov's formula 
\cite{Nekrasov} of the instanton sum for the gauge theories 
in the $\Omega$ background \cite{Losev-Marshakov-Nekrasov}.  
This statistical model 
has been extended \cite{Marshakov-Nekrasov} to investigate  
the integrability of correlation functions of 
single-traced chiral observables. 
Such an extension of the Seiberg-Witten theory  
also becomes attractive to understand $4d$ $\mathcal{N}=1$ 
supersymmetric gauge theories by providing a powerful tool 
\cite{Itoyama}.

The so-called melting crystal model provides 
a similar statistical model for $5d$ $\mathcal{N}=1$ 
supersymmetric gauge theories \cite{MNTT1, MNTT2, MNNT} 
and $A$-model topological strings \cite{Crystal}.
This is a statistical model of {\it random plane partitions}, 
also interpreted  as a $q$-deformation of random partitions
\cite{Nekrasov-Okounkov, Marino at al, Iqbal et al 2008}, 
that stems from Nekrasov's formula \cite{Nekrasov} 
for $5d$ supersymmetric gauge theories in the $\Omega$ background.

Recently, an extension of this melting crystal model 
with external potentials was proposed, and 
its integrable structure was elucidated 
\cite{Nakatsu-Takasaki, Nakatsu-Noma-Takasaki}.  
It was argued therein that the external potentials 
are related to loop operators of 
a $5d$ $\mathcal{N}=1$ supersymmetric Yang-Mills theory (SYM).  
The goal of this paper is to present the detail of 
this result along with some implications.  
In particular, we study the thermodynamic limit 
of the partition function of this model, and 
solve a Riemann-Hilbert problem that determines 
the limit shape of the main diagonal slice of 
random plane partitions in the presence of 
external potentials.  We can thus eventually 
identify a relevant complex curve and the associated 
Seiberg-Witten differential.

This paper is organized as follows.  
Section 2 starts with a brief review about $5d$ $\mathcal{N}=1$ SYM 
in the $\Omega$ background. 
We introduce loop operators $\mathcal{O}_k$ $(k=1,2,\cdots)$ 
of this theory. 
Computation of correlation functions 
among these operators  is presented  
by using the localization technique. 
Generating function of the correlation functions of $U(1)$ theory 
is expressed as a statistical sum over partitions 
and reproduces the partition function of 
the aforementioned melting crystal model. 
For an application of the localization theorem, 
we use equivariant descent equations 
of the loop operators, 
which is proved in Appendix B, 
and a related $T^2$-action on the gauge theory 
is argued in Appendix A. 
Section 3 is concerned with 
a common integrable structure of 
$5d$ $\mathcal{N}=1$ SYM in the $\Omega$ background 
and melting crystal model. 
The loop operators are converted to the external potentials 
of the melting crystal model. 
This eventually shows that, 
by regarding the coupling constants $t=(t_1,t_2,\cdots)$ of 
the loop operators as a series of time variables, 
the generating function becomes a $\tau$ function of $1$-Toda 
hierarchy. 
In Section 4, 
we introduce an energy functional. 
This functional is a quadratic form, 
and obtained from the logarithm of the statistical weight 
in the partition function. 
The formula is stated in a general setting and is proved in Appendix C. 
By using the formula,  
we argue the thermodynamic limit of the melting crystal model 
or the $q$-deformed random partitions. 
In Section 5, 
the thermodynamic limit is reformulated  
as a Riemann-Hilbert problem 
to obtain an analytic function on 
$\mathbb{C}^*-I$, 
where 
$\mathbb{C}^*$ is a cylinder and 
$I$ is an interval in the real axis, 
and satisfies suitable conditions. 
In Section 6, 
we solve the Riemann-Hilbert problem. 
The relevant complex curve and the associated 
Seiberg-Witten differential are presented.  
The vev's of the loop operators 
are particularly expressed as residue integrals 
of the Seiberg-Witten differential. 
The solution of the Riemann-Hilbert problem can be 
further analyzed by applying the classical Jensen formula 
in complex analysis. 
The case of the single coupling constant 
$t=(t_1,0,0,\cdots)$ is described as an example. 
The related computation is attached to Appendix D. 
Section 7 is devoted to conclusion and discussion.

\section{Loop operators of 
$5d$ $\mathcal{N}=1$ SYM in $\Omega$ background}

We first consider 
an ordinary $5d$ $\mathcal{N}=1$ SYM on $\mathbb{R}^4\times S^1$. 
Let $E$ be the $SU(N)$-bundle on $\mathbb{R}^4$ with $c_2(E)=n \geq 0$. 
A gauge bundle of the theory is the $SU(N)$-bundle $\pi^*E$ on 
$\mathbb{R}^4\times S^1$ pulled back from $\mathbb{R}^4$, 
where $\pi$ is the projection from 
$\mathbb{R}^4\times S^1$ to $\mathbb{R}^4$.  
All the fields in the vector multiplet 
are set to be periodic along $S^1$. 
Among them, 
the bosonic ingredients are a $5d$ gauge potential $A_M(x,t)dx^M$ 
and a scalar field $\varphi(x,t)$ 
taking the value in $su(N)$, 
where the coordinates $x^M$ represent 
coordinates $x=(x^1,x^2,x^3,x^4)$ of $\mathbb{R}^4$ 
and a periodic coordinate $t$ of $S^1$. 
These bosonic fields describe a $5d$ Yang-Mills-Higgs system.  
The gauge potential can be separated into two parts 
$A_{\mu}(x,t)dx^{\mu}$ and $A_t(x,t)dt$, 
respectively the components 
of the $\mathbb{R}^4$- and the $S^1$-directions. 
Let $\mathcal{A}_E$ be the infinite dimensional affine space 
consisting of all the gauge potentials on $E$. 
The $4d$ component $A_{\mu}(x,t)dx^{\mu}$ 
describes a loop $A(t)$ in $\mathcal{A}_E$, 
where the loop is parametrized by 
the periodic coordinate of the fifth-dimensional circle. 
As for $A_t(x,t)$, together with $\varphi(x,t)$, 
the combination $A_t+i\varphi$ describes a loop $\phi(t)$ 
in $\Omega^0(\mathbb{R}^4,\mbox{ad}E \otimes \mathbb{C})$, 
which is the space of all the sections of 
$\mbox{ad}E \otimes \mathbb{C}$, 
where ad$E$ is the adjoint bundle on $\mathbb{R}^4$ with fibre $su(N)$. 
Taking account of the periodicity,  
the same argument is applicable to the gauginos as well. 
The vector multiplet thereby describes 
a loop in the configuration space of the $4d$ theory. 
By using such a loop, 
we may describe $5d$ $\mathcal{N}=1$ SYM. 
In the case of the Yang-Mills-Higgs system, 
the loop $A(t)$ gives 
a family of covariant differentials on $E$ as  $d_{A(t)}=d+A(t)$.   
For the loop $\phi(t)$, 
since it involves $A_t(x,t)$, 
we conveniently 
introduce a differential operator $\mathcal{H}(t)$ by 
\begin{eqnarray}
\mathcal{H}(t)\equiv 
\frac{d}{dt}+\phi(t)\,. 
\label{H(t)}
\end{eqnarray}

\subsection{$5d$ $\mathcal{N}=1$ SYM in $\Omega$ background}

Via the standard dimensional reductions,
$6d$ $\mathcal{N}=1$ SYM gives 
lower dimensional Yang-Mills theories with $8$ supercharges, 
including the above theory. 
Furthermore, 
the dimensional reductions 
in the $\Omega$ background 
provide much powerful tools 
to understand these theories 
\cite{Losev-Marshakov-Nekrasov}.
The $\Omega$ background is 
a $6d$ gravitational background 
on $\mathbb{R}^4 \times T^2$ 
described by a metric of the form: 
${\displaystyle 
ds^2=
\sum_{\mu=1}^4
\bigl(
dx^{\mu}-\sum_{a=5,6}V_a^{\mu}dx^a
\bigr)^2
+\sum_{a=5,6}
\bigl(
dx^a
\bigr)^2\,, }$
where two vectors $V_5^{\mu},V_6^{\mu}$ 
generate rotations on two-planes $(x^1,x^2)$ and $(x^3,x^4)$ 
in $\mathbb{R}^4$.  
By letting  
$V_1=
x^2\frac{\partial}{\partial x^1}
-x^1\frac{\partial}{\partial x^2}$ 
and 
$V_2=
x^4\frac{\partial}{\partial x^3}
-x^3\frac{\partial}{\partial x^4}$, 
they are respectively 
the real part and the imaginary part of the combination   
\begin{eqnarray}
V_{\epsilon_1,\epsilon_2}
&\equiv& \epsilon_1V_1+\epsilon_2V_2\,,  
\hspace{6mm}\epsilon_1,\epsilon_2 \in \mathbb{C}.
\label{V_epsilon}
\end{eqnarray}
The above combination is expressed in component as 
$V_{\epsilon_1,\epsilon_2}
=\Omega^\mu_{~\nu} x^\nu\frac{\partial}{\partial x^\mu}$.

To see the dimensional reduction in the $\Omega$-background, 
let us first consider the bosonic part of the $5d$ SYM. 
The corresponding Yang-Mills-Higgs system is modified 
from the previous one.  
However, 
the system is eventually controlled by 
replacing $\mathcal{H}(t)$ with  
\begin{eqnarray}
\mathcal{H}_{\epsilon_1,\epsilon_2}(t)
\equiv 
\mathcal{H}(t)+\mathcal{K}_{\epsilon_1,\epsilon_2}(t)\,.
\label{H_epsilon}
\end{eqnarray}
Here $\mathcal{K}_{\epsilon_1,\epsilon_2}(t)$ 
is an another differential operator given by 
\cite{student}
\begin{eqnarray}
\mathcal{K}_{\epsilon_1,\epsilon_2}(t)
\equiv 
V_{\epsilon_1,\epsilon_2}^\mu \partial_{A(t)\,\mu}
+
\frac{1}{2}\Omega^{\mu \nu}\mathcal{J}_{\mu \nu}\,, 
\label{K_epsilon}
\end{eqnarray}
where 
$\mathcal{J}_{\mu \nu}$ denote 
the $SO(4)$ Lorentz generators of the system. 
The above operator generates a $T^2$-action  
by taking the commutators with $d_{A(t)}$ and $\mathcal{H}(t)$. 
For instance, we have 
\begin{eqnarray} 
\bigl[\,
d_{A(t)},\, 
\mathcal{K}_{\epsilon_1,\epsilon_2}(t)\,
\bigr]
=-\iota_{V_{\epsilon_1,\epsilon_2}}F_{A(t)}\,, 
\label{torus action on A}
\end{eqnarray} 
where the right-hand side means $-1$ times 
the contraction of the curvature two-form 
$F_{A(t)}=dA(t)+A(t)\wedge A(t)$ 
with the vector field $V_{\epsilon_1,\epsilon_2}$. 
As is argued in Appendix \ref{torus_action_on_A}, 
the infinitesimal rotation 
$\delta x^\mu=-V^{\mu}_{\epsilon_1,\epsilon_2}$ 
can generate a $T^2$-action on $\mathcal{A}_E$. 
The right-hand side of (\ref{torus action on A}) 
is precisely the infinitesimal form (\ref{t_epsilon_A})
of the $T^2$-action.

The supercharges $Q_{\alpha a}$ and $\bar{Q}^{\dot{\alpha}}_{a}$ 
are realized in a way different from the case 
of $\epsilon_1=\epsilon_2=0$.  
Note that we use the $4d$ notation 
such that $\alpha,\dot{\alpha}$ and $a$ 
denote the indexes of the Lorentz group $SU(2)_L\times SU(2)_R$ 
and the R-symmetry $SU(2)_I$. 
According to the argument \cite{Nekrasov_1996}, 
we may interpret the $5d$ SYM  
as a topological field theory \cite{Witten}. 
Actually, 
by regarding the diagonal $SU(2)$ of $SU(2)_R\times SU(2)_I$ as 
a new $SU(2)_R$, we can extract a supercharge that behaves as 
a scalar under the new Lorentz symmetry. 
We write the scalar supercharge as $Q_{\epsilon_1,\epsilon_2}$. 
The gaugino acquires a natural interpretation  
as differential forms,  
$\eta(x,t),\psi_{\mu}(x,t)$ and $\xi_{\mu \nu}(x,t)$. 
These give 
fermionic loops, $\eta(t), \psi(t)$ and $\xi(t)$.  
The main part of the $Q$-transformation can be read as 
\begin{eqnarray}
&&
Q_{\epsilon_1,\epsilon_2}A(t)=\psi(t)\,, 
\hspace{7mm}
Q_{\epsilon_1,\epsilon_2}\psi(t)=
\bigl[\,
d_{A(t)},\, 
\mathcal{H}_{\epsilon_1,\epsilon_2}(t)\,
\bigr]\,,
\label{Q transform (A, psi)}
\\
&&
Q_{\epsilon_1,\epsilon_2}
\mathcal{H}_{\epsilon_1,\epsilon_2}(t)=0\,,
\label{Q transform H}
\end{eqnarray}
where $\psi(t)$ is a fermionic loop in 
$\Omega^1(\mathbb{R}^4,\mbox{ad}E)$.

The action of the $5d$ SYM can be now written 
in a $Q$-exact form as 
\begin{eqnarray}
S_{5d\,SYM}^{\epsilon_1,\epsilon_2}= 
\frac{4\pi^2n}{g^2}\int_{S^1} dt 
+\int_{S_1}dt 
\Bigl\{ Q_{\epsilon_1,\epsilon_2}, \mathcal{W}(t)\Bigr\}\,. 
\label{5d SYM_epsilon}
\end{eqnarray}
The above action turns out to 
be a BRST gauge fixed action 
of the topological term $\frac{4\pi^2n}{g^2}\int_{S^1} dt$, 
where $Q_{\epsilon_1,\epsilon_2}$ plays a role of the BRST charge.  
In this formulation, 
the original vector multiplet is splitted to 
several $Q$-multiplets, 
including $A(t),\psi(t)$ as the doublet and  
$\mathcal{H}_{\epsilon_1,\epsilon_2}(t)$ as the singlet. 
We may integrate out all the multiplets except the two. 
The integrations impose constraints on the two multiplets. 
As for the doublet, 
the constraints turn to involve three equations;  
$F_{A(t)}^{(+)}=0$,
$\Big(d_{A(t)}\psi(t)\Bigr)^{(+)}=0$, 
where the symbol $(+)$ means self-dual part of two-forms 
on $\mathbb{R}^4$,  
and 
$d_{A(t)}^*\psi(t)=0$. 
The first equation is the anti-self dual equation for 
gauge potentials on $E$.  
The second equation is regarded as a linearization of the first. 
For the singlet, 
the constraint leads to the equation
\begin{eqnarray}
\square_{A(t)}\phi(t)+
2\psi(t)^2+
d_{A(t)}^*
\left( 
\iota_{V_{\epsilon_1,\epsilon_2}}F_{A(t)}
+\frac{dA(t)}{dt}
\right)=0\,, 
\label{constraint on phi(t)}
\end{eqnarray}
where $\psi(t)^2=\psi_\mu(t)\psi^\mu(t)$. 
In the above, 
$d_{A(t)}^*$ is the formal adjoint of $d_{A(t)}$ 
and $\square_{A(t)}=(-1)d_{A(t)}^*d_{A(t)}$ is the scalar Laplacian.

\subsection{Loop operators and their correlation functions}

Taking account of the relation $\phi(x,t)=A_t(x,t)+i\varphi(x,t)$, 
the following path-ordered integral 
becomes an analogue of holonomy of the 5$d$ gauge potential: 
\begin{eqnarray}
W^{(0)}(x;t_1,t_2)=
\mbox{P}\exp
\Bigl(
-\int_{t_2}^{t_1}dt\phi(x,t)
\Bigr)\,, 
\label{W_(0)}
\end{eqnarray}
where the symbol means path-ordered integration, 
more precisely, 
it is defined by the differential equation  
\begin{eqnarray}
(\frac{d}{dt_1}+\phi(x,t_1))W^{(0)}(x;t_1,t_2)=0\,, 
\hspace{5mm}
W^{(0)}(x;t_2,t_2)=1\,. 
\label{def O_(0)}
\end{eqnarray} 
The trace of the holonomy along the circle 
defines a loop operator by 
\begin{eqnarray}
\mathcal{O}^{(0)}(x)= 
\mbox{Tr}\, W^{(0)}(x; R,0)\,, 
\label{O_(0)}
\end{eqnarray}
where $R$ is the circumference of $S^1$. 
The above operator is an analogue of 
the Wilson loop operator along the circle. 
Unlike the case of $\epsilon_1=\epsilon_2=0$
\cite{Baulieu-Losev-Nekrasov}, 
it is not $Q$-closed except at $x=0$. 
To see this, note that the $Q$-transformations 
(\ref{Q transform (A, psi)}) and (\ref{Q transform H}) 
imply    
$Q_{\epsilon_1,\epsilon_2}\phi(t)=
-\iota_{V_{\epsilon_1,\epsilon_2}}\psi(t)$. 
By using this,  
we obtain the transformation 
\begin{eqnarray}
Q_{\epsilon_1,\epsilon_2}\mathcal{O}^{(0)}(x)
=
\int_0^{R}dt
\mbox{Tr}\Bigl\{\,W^{(0)}(x;\,R,t)\,
\iota_{V_{\epsilon_1,\epsilon_2}}\psi(x,t)\,
W^{(0)}(x;\,t,0)\Bigr\}\,. 
\label{QO_(0)}
\end{eqnarray}
Since the right-hand side of the above formula vanishes only at $x=0$,  
this means that $\mathcal{O}^{(0)}(x)$ becomes $Q$-closed only at $x=0$.

The above property may be explained 
in terms of the equivariant de Rham theory. 
To see this, 
let us first generalize the path-ordered integral (\ref{W_(0)}) 
by exponentiating the combination 
$F_{A(t)}-\psi(t)+\phi(t)$ in place of $\phi(t)$ as 
\begin{eqnarray}
W(x;\,t_1,t_2)
=\mbox{P}
\exp
\Bigl\{
-\int_{t_2}^{t_1}dt \big(F_{A(t)}-\psi(t)+\phi(t)\big)(x)
\Bigr\}\,, 
\label{W}
\end{eqnarray}
where the right-hand side is the solution of 
the differential equation 
\begin{eqnarray}
\frac{dW(x;t_1,t_2)}{dt_1}+
\Bigl(
F_{A(t_1)}(x)-\psi(x,t_1)+\phi(x,t_1)
\Bigr)
\wedge 
W(x;t_1,t_2)&=&0, 
\nonumber \\
W(x;t_2,t_2)&=&0\,. 
\end{eqnarray}
This particularly 
means that $W$ has several components, 
according to degrees of differential forms on $\mathbb{R}^4$, 
as $W=W^{(0)}+W^{(1)}+\cdots+W^{(4)}$, 
where the indexes denote the degrees.
We then generalize the loop operator (\ref{O_(0)}) as 
\begin{eqnarray}
\mathcal{O}(x)=
\mbox{Tr}\, W(x; R,0)\,.
\label{O}
\end{eqnarray}
In parallel with $W$, 
we have the decomposition 
$\mathcal{O}=\mathcal{O}^{(0)}
+\mathcal{O}^{(1)}+\cdots+\mathcal{O}^{(4)}$.  
Explicitly, 
$\mathcal{O}^{(i)}$ are operators of the following form: 
\begin{eqnarray}
\mathcal{O}^{(1)}(x)&=&
\int_0^{R}dt
\mbox{Tr}\Bigl\{\,W^{(0)}(x;R,t)
\psi(x,t)
W^{(0)}(x;t,0)\Bigr\}\,, 
\nonumber \\
\mathcal{O}^{(2)}(x)&=&
\int_0^{R}dt
\mbox{Tr}\Bigl\{\,W^{(0)}(x;R,t)
(-)F_{A(t)}(x)
W^{(0)}(x;t,0)\Bigr\}
\nonumber \\
&&
\hspace{-3mm}
+\int_0^{R}dt_1
\int_0^{t_1}dt_2
\mbox{Tr}\Bigl\{\,W^{(0)}(x;R,t_1)
\psi(x,t_1)
W^{(0)}(x;t_1,t_2)
\psi(x,t_2)
W^{(0)}(x;t_2,0)
\Bigr\}\,, 
\nonumber \\
\vdots 
&&
\label{O^(1,2)}
\end{eqnarray}

Eq.(\ref{QO_(0)}) can be now expressed as 
$Q_{\epsilon_1,\epsilon_2}\mathcal{O}^{(0)}
=\iota_{V_{\epsilon_1,\epsilon_2}}\mathcal{O}^{(1)}$, 
which is actually the first equation 
among a series of the equations that $\mathcal{O}^{(i)}$ obey. 
Such equations eventually show up 
by expanding component-wise the identity
\begin{eqnarray}
(d_{\epsilon_1,\epsilon_2}+Q_{\epsilon_1,\epsilon_2})\mathcal{O}(x)=0\,, 
\label{formula of O}
\end{eqnarray}
where $d_{\epsilon_1,\epsilon_2}\equiv 
d-\iota_{V_{\epsilon_1,\epsilon_2}}$
is the $T^2$-equivariant differential on $\mathbb{R}^4$. 
The above identity implies in components 
$T^2$-equivariant descent equations of the form
\begin{eqnarray}
d \mathcal{O}^{(i-1)}(x)
+Q_{\epsilon_1, \epsilon_2}\mathcal{O}^{(i)}(x)
-\iota_{V_{\epsilon_1, \epsilon_2}}\mathcal{O}^{(i+1)}(x)
=0\,,
\hspace{10mm} 
0 \leq i \leq 4\,,
\label{formula_of_O_in_components}
\end{eqnarray}
where $\mathcal{O}^{(-1)}=\mathcal{O}^{(5)}=0$. 
We provide a proof of the formula (\ref{formula of O}) 
in Appendix \ref{Proof:formula of O}.

We can also consider the loop operators 
encircling the circle many times. 
Correspondingly, 
we introduce 
\begin{eqnarray}
\mathcal{O}_k(x)\equiv 
\mbox{Tr}\,W(x;kR,0)\,, 
\hspace{6mm}
k=1,2,\cdots .
\label{O_k} 
\end{eqnarray}
These satisfy 
\begin{eqnarray}
(d_{\epsilon_1,\epsilon_2}+Q_{\epsilon_1,\epsilon_2})\mathcal{O}_k(x)=0\,. 
\label{formula of O_k}
\end{eqnarray}

Let us examine the correlation functions 
$\langle\, 
\prod_{a}\int_{\mathbb{R}^4}\mathcal{O}_{k_a}\,
\rangle^{\epsilon_1,\epsilon_2}$. 
Since the integral
$\int_{\mathbb{R}^4}\mathcal{O}_k=\int_{\mathbb{R}^4}\mathcal{O}_k^{(4)}$
is $Q$-closed by virtue of the formula (\ref{formula of O_k}), 
the correlation function has an interpretation 
in the topological field theory. 
In particular, 
we may compute the correlation functions    
by a supersymmetric quantum mechanics (SQM) 
which is substantially equivalent to 
the $5d$ SYM as the topological field theory. 
Such a SQM turns out to be 
an equivariant SQM on $\tilde{\mathcal{M}}_n$   
\cite{Nekrasov}, 
where $\tilde{\mathcal{M}}_n$ 
is the moduli space of the framed $n$ instantons.

We may regard $\tilde{\mathcal{M}}_n$ 
as a Riemannian manifold. 
Actually, 
by taking a local gauge slice of 
the anti-self dual gauge potentials, 
the metric is induced from $\mathcal{A}_{E}$, 
where $\mathcal{A}_{E}$ is endowed with 
a gauge invariant metric of the form,   
$G(\delta A,\delta' A)=
\int_{\mathbb{R}^4}\mbox{Tr}\delta A(x)\wedge * \delta' A(x)$. 
Take a loop $m(t)=(m^I(t))$, where $m^I$ denote 
local coordinates of $\tilde{\mathcal{M}}_n$.  
We also introduce its fermionic partner 
$\chi(t)=(\chi^I(t))$. 
Both are parametrized by the fifth-dimensional circle. 
The $Q$-transformation (\ref{Q transform (A, psi)}) 
yields a transformation between them. 
To describe the transformation, 
note that the infinitesimal variation 
$\delta A=-\iota_{V_{\epsilon_1,\epsilon_2}}F_A$ 
preserves the anti-self dual equation 
$F_A^{(+)}=0$, 
thus it gives a vector field on $\tilde{\mathcal{M}}_n$, 
which we denote by  
$\mathcal{V}_{\epsilon_1,\epsilon_2}$. 
It is actually a Killing vector. 
This apparently follows  since the aforementioned 
variation preserves the gauge invariant metric $G$. 
The $Q$-transformation (\ref{Q transform (A, psi)}) 
is eventually converted to the transformation 
\begin{eqnarray}
Q_{\epsilon_1,\epsilon_2}m(t)=\chi(t)\,, 
\hspace{7mm}
Q_{\epsilon_1,\epsilon_2}\chi(t)=
-\frac{dm(t)}{dt}+\mathcal{V}_{\epsilon_1,\epsilon_2}(m(t))\,.
\label{Q transform (m,chi)}
\end{eqnarray}

The supersymmetric Yang-Mills is described 
effectively in terms of $m(t)$ and $\chi(t)$,
consequently reduces to 
a quantum mechanical system on $\tilde{\mathcal{M}}_n$. 
Actually, 
the corresponding action can be obtained from 
(\ref{5d SYM_epsilon}) 
by integrating out the irrelevant fields. 
This yields the action  
\begin{eqnarray}
S_{eff}^{\epsilon_1,\epsilon_2}= 
\frac{4\pi^2n}{g^2}\int_{S^1} dt 
+\int_{S_1}dt 
\left\{ Q_{\epsilon_1,\epsilon_2}, 
\frac{1}{2}G_{IJ}(m(t))\chi^I(t)\frac{dm^J(t)}{dt}
\right\}\,,  
\label{5d SYM_epsilon_effective}
\end{eqnarray}
where $G_{IJ}$ is the Riemannian metric on $\tilde{\mathcal{M}}_n$.  
The above action together with 
the supersymmetry (\ref{Q transform (m,chi)}) is familiar 
in the physical proof \cite{Niemi-Tirkkonens} 
of the equivariant index formula for Dirac operator. 
In particular, 
the partition function becomes eventually 
the $T^2$-equivariant index 
for the Dirac operator on $\tilde{\mathcal{M}}_n$ 
\cite{Nekrasov}.

The combination $F_{A(t)}-\psi(t)+\phi(t)$ in (\ref{W}) 
can be identified with a loop space analogue of 
the $T^2$-equivariant curvature $\mathcal{F}_{\epsilon_1,\epsilon_2}$ 
of the universal connection 
\cite{Donaldoson-Kronheimer}, 
where the universal bundle becomes equivariant by 
the $T^2$-action on $\mathcal{A}_E \times \mathbb{R}^4$.  
In the computation of the correlation function, 
by virtue of the supersymmetry (\ref{Q transform (m,chi)}),  
only the constant modes $m_0,\chi_0$ contribute to 
the observables. 
For such constant modes, 
the above combination is precisely identified 
with the equivariant curvature $\mathcal{F}_{\epsilon_1,\epsilon_2}$
\cite{Losev-Marshakov-Nekrasov}. 
This means that $\mathcal{O}_k(x)$ substantially truncates 
to the $T^2$-equivariant Chern character 
$\mbox{Tr} \, e^{-kR\, \mathcal{F}_{\epsilon_1,\epsilon_2}}$.
Thus we obtain the following finite dimensional 
integral representation of the correlation functions: 
\begin{eqnarray}
\Big\langle 
\,\prod_{a}\int_{\mathbb{R}^4}\mathcal{O}_{k_a}\,
\Big \rangle_{n-instanton}^{\epsilon_1,\epsilon_2}
=
\frac{1}{(2\pi i R)^{\frac{\dim \tilde{\mathcal{M}}_n}{2}}}
\int_{\tilde{\mathcal{M}}_n}
\hat{A}_{T^2}(R\,{\bf t}_{\epsilon_1,\epsilon_2},
\,\tilde{\mathcal{M}}_n)\,
\prod_{a}
\int_{\mathbb{R}^4} 
\mbox{Tr} \, 
e^{-k_aR \, \mathcal{F}_{\epsilon_1,\epsilon_2}}\,. 
\label{correlator of O_a}
\end{eqnarray}
where $\hat{A}_{T^2}(\cdot\,,\tilde{\mathcal{M}}_n)$ is 
the $T^2$-equivariant $\hat{A}$-genus of the tangent bundle 
of $\tilde{\mathcal{M}}_n$, 
and 
$\bf{t}_{\epsilon_1,\epsilon_2}$ 
is a generator of $T^2$ that gives the Killing vector 
$\mathcal{V}_{\epsilon_1,\epsilon_2}$.

Introducing coupling constants $t=(t_1,t_2,\cdots)$,  
the generating function of the correlation functions 
is given by 
$\mathcal{Z}_{\epsilon_1,\epsilon_2}(t)
=\left \langle 
e^{\sum_{k}t_k \int_{\mathbb{R}^4}\mathcal{O}_k} 
\right \rangle^{\epsilon_1,\epsilon_2}$. 
Since $n$-instanton 
contributes with the weight $(R\Lambda)^{2nN}$, 
where $\Lambda$ is the dynamical scale,  
letting $Q=(R\Lambda)^2$, 
we can express the generating function as 
\begin{eqnarray}
\mathcal{Z}_{\epsilon_1,\epsilon_2}(t)
=
\sum_{n=0}^{\infty}
Q^{nN}
\left \langle 
e^{\sum_{k}t_k \int_{\mathbb{R}^4}\mathcal{O}_k} 
\right \rangle_{n-instanton}^{\epsilon_1,\epsilon_2}  \,. 
\label{generating function SU(N)}
\end{eqnarray}

\subsection{Application of localization technique}

The right-hand side of 
the formula (\ref{correlator of O_a})
is eventually replaced with a statistical sum over partitions. 
To see their appearance, 
note that the integration 
localizes to the fixed points of the $T^2$-action. 
However, 
the fixed points in $\tilde{\mathcal{M}}_n$ 
are small instanton singularities  
since the variation 
$\delta A= -\iota_{V_{\epsilon_1,\epsilon_2}}F_A$ 
vanishes there. 
These singularities can be resolved by instantons 
on a non-commutative $\mathbb{R}^4$. 
By using such a regularization via the non-commutativity, 
the fixed points get isolated, 
so that they are eventually labelled by using partitions 
\cite{Nakajima}.

A partition $\lambda=(\lambda_1,\lambda_2,\cdots)$ is 
a sequence of non-negative integers 
satisfying $\lambda_i \geq \lambda_{i+1}$ 
for all $i \geq 1$. 
Partitions are identified with the Young diagrams 
in the standard manner. 
The size is defined by $|\lambda|=\sum_{i \geq 1}\lambda_i$, 
which is the total number of boxes of the diagram.

Let us examine the formula (\ref{correlator of O_a}) 
for the $U(1)$ theory. 
The relevant computation of the localization 
can be found in \cite{Nakajima,Nakajima-Yoshioka-lec}. 
We truncate $\epsilon_{1,2}$ as 
$-\epsilon_{1}=\epsilon_2=i\hbar$, 
where $\hbar$ is a positive real parameter. 
Consequently, 
the formula becomes a $q$-series by the truncation, 
where $q=e^{-R\hbar}$.  
The fixed points in $\tilde{\mathcal{M}}_n$ 
are labelled by partitions of $n$. 
The equivariant $\hat{A}$-genus takes the following form 
at the partition $\lambda$ of $n$: 
\begin{eqnarray}
(2\pi iR)^{-2n}
\left. 
\hat{A}_{T^2}(R\,{\bf t}_{-i\hbar,i\hbar}, 
\,\tilde{\mathcal{M}}_n)
\right |_{\lambda}
=
(-)^n
\Bigl(\frac{\hbar}{2\pi}\Bigr)^{2n}
\prod_{s \in \lambda}
\left\{
\frac{h(s)}{q^{-\frac{h(s)}{2}}-q^{\frac{h(s)}{2}}}
\right\}^2\,, 
\label{U(1) Dirac_index_product}
\end{eqnarray}
where $h(s)$ denotes the hook length of the box $s$ of 
the Young diagram $\lambda$. 
The right-hand side of the above formula can be expressed 
by using a Schur function as 
\begin{eqnarray}
(2\pi iR)^{-2n}
\left. 
\hat{A}_{T^2}(R\,{\bf t}_{-i\hbar,i\hbar}, 
\,\tilde{\mathcal{M}}_n)
\right |_{\lambda}
=
(-)^{n}
\Bigl(\frac{\hbar}{2\pi}\Bigr)^{2n}
\Bigl(\prod_{s \in \lambda} h(s)\Bigr)^2 
q^{\frac{\kappa(\lambda)}{2}}
s_{\lambda}(q^{\rho})^2\,, 
\label{U(1) Dirac_index}
\end{eqnarray}
where $s_{\lambda}(q^{\rho})$ is the Schur function 
$s_{\lambda}(x_1,x_2,\cdots)$ specialized to $x_i=q^{i-\frac{1}{2}}$, 
and 
$\kappa(\lambda)=2\sum_{(i,j) \in \lambda}(j-i)$. 
To obtain (\ref{U(1) Dirac_index}), 
we have made use of the $q$-hook formula \cite{Macdonald}
\begin{eqnarray}
s_{\lambda}(q^{\rho})=
q^{n(\lambda)+\frac{|\lambda|}{2}}
\prod_{s \in \lambda}
\Bigl(1-q^{h(s)}\Bigr)^{-1}\,, 
\label{q-hook_formula}
\end{eqnarray}
where $n(\lambda)=\sum_{i \geq 1}(i-1)\lambda_i$. 
Actually, taking account of the formula 
$2n(\lambda)+|\lambda|=-\kappa(\lambda)/2+\sum_{s \in \lambda}h(s)$, 
the $q$-hook formula gives  
$\prod_{s \in \lambda}
\Bigl(
q^{-\frac{h(s)}{2}}-q^{\frac{h(s)}{2}}
\Bigr)^{-1}
=
q^{\frac{\kappa(\lambda)}{4}}s_{\lambda}(q^{\rho})$.
By plugging this into the right-hand side of 
(\ref{U(1) Dirac_index_product}), 
we obtain (\ref{U(1) Dirac_index}).

Similarly,  
the fixed points in $\tilde{\mathcal{M}}_n \times \mathbb{R}^4$ 
are given as $(\lambda,0)$, 
where $\lambda$ is a partition of $n$ and 
$0$ is the origin of $\mathbb{R}^4$.   
The equivariant Chern character takes the form  
$
\left.
\mbox{Tr}\,e^{-kR \mathcal{F}_{-i\hbar,i\hbar}}
\right|_{(\lambda,0)}
=\mathcal{O}_k(\lambda)
$, 
where $\mathcal{O}_k(\lambda)$ is given by 
\begin{eqnarray}
\mathcal{O}_{k}(\lambda)
=
(1-q^{-k})
\sum_{i=1}^{\infty}
\Bigl\{
q^{k(\lambda_i-i+1)}-q^{k(-i+1)}
\Bigr\}
+1\,. 
\label{O_l(lambda)}
\end{eqnarray}
The above functions have been exploited in 
\cite{Marshakov-Nekrasov, Kanno-Moriyama} 
from the $4d$ gauge theory viewpoint, with $q$ or $q^k$ 
being replaced by a generating spectral parameter.

By the localization formula \cite{Berline_Getzler_Vergne}, 
integration in the right-hand side of (\ref{correlator of O_a}) 
reduces to a summation over contributions from 
the isolated fixed points of the $T^2$-action on 
$\tilde{\mathcal{M}}_n \times 
(\mathbb{R}^4\times \cdots \times \mathbb{R}^4)$. 
The contribution from the fixed point consists of two factors. 
One is the value of the integrand at the fixed point, 
and the other is the Jacobian factor in a change of coordinates 
in a neighborhood of the fixed point.  
By multiplying these two factors and 
summing up them over the fixed points, 
we obtain the right-hand side of (\ref{correlator of O_a}). 
The fixed points are of the form $(\lambda,(0,\cdots,0))
\in \tilde{\mathcal{M}}_n \times 
(\mathbb{R}^4\times \cdots \times \mathbb{R}^4)$, 
where $\lambda$ is a partition of $n$ and 
$(0,\cdots,0)$ is the origin of  
$\mathbb{R}^4\times \cdots \times \mathbb{R}^4$. 
The values of the integrand at the fixed points 
can be expressed by using the formulas 
(\ref{U(1) Dirac_index}) and (\ref{O_l(lambda)}). 
The Jacobian factor can be further factorized by 
the product structure of 
$\tilde{\mathcal{M}}_n$ and $\mathbb{R}^4$'s.  
The Jacobian factor originating 
in the change of coordinates of $\tilde{\mathcal{M}}_n$ 
in a neighborhood of $\lambda$ takes the form 
$\bigl(2\pi/\hbar\bigr)^{2n}(\prod_{s \in \lambda}h(s))^{-2}$,  
while 
the equivariant volume of 
$\mathbb{R}^4\times \cdots \times \mathbb{R}^4$ 
is given by the products of $\hbar^{-2}$
for each $0 \in \mathbb{R}^4$.
Therefore, 
the formula (\ref{correlator of O_a}) becomes 
eventually the statistical sum over partitions given by 
\begin{eqnarray}
\Big\langle 
\,\prod_{a}\int_{\mathbb{R}^4}\mathcal{O}_{k_a}\,
\Big \rangle_{n-instanton}^{-i\hbar,i\hbar}
=
(-)^n
\sum_{\lambda\,:\,\,partition\,of\,\,n}
q^{\frac{\kappa(\lambda)}{2}}
s_{\lambda}(q^{\rho})^2
\prod_{a}
\hbar^{-2}
\mathcal{O}_{k_a}(\lambda)\,,
\label{correlator of O_a_U(1)}
\end{eqnarray}
where the summation is taken over partitions of $n$.

Although we have not taken into account, 
the Chern-Simon term can be added 
to a $5d$ gauge theory,   
with the coupling constant being quantized, 
in particular,  
for the $U(1)$ theory, 
$m=0,\pm 1$. 
It modifies  
the right-hand side of (\ref{correlator of O_a_U(1)}) 
by giving a contribution of the form 
$(-)^{m|\lambda|}q^{-\frac{m\kappa(\lambda)}{2}}$,  
for each $\lambda$ \cite{MNTT1}.  
Hereafter, 
we consider the case of 
the $U(1)$ theory having  
the Chern-Simon coupling, $m=1$. 
The corresponding generating function 
(\ref{generating function SU(N)}) becomes  
\begin{eqnarray}
\mathcal{Z}^{U(1)}_{-i\hbar,i\hbar}(t)
=
\sum_{\lambda} 
Q^{|\lambda|} 
s_{\lambda}(q^{\rho})^2
e^{\hbar^{-2}\sum_{k=1}^{\infty}t_k \mathcal{O}_k(\lambda)}\,, 
\label{Z_U(1)}
\end{eqnarray}
where the summation is taken 
over all the partitions.

\section{Integrability of  $5d$ $\mathcal{N}=1$ SYM 
in $\Omega$ background}

We can regard the generating function (\ref{Z_U(1)}) 
as a $q$-deformed random partition. 
To see this,  
note that the $4d$ limit $R \rightarrow 0$ makes 
$q=e^{-R\hbar} \rightarrow 1$, 
therefore, 
by using (\ref{q-hook_formula}), 
the Boltzmann weight 
$Q^{|\lambda|} s_{\lambda}(q^{\rho})^2$ 
in (\ref{Z_U(1)}) 
takes at this limit,  
the form    
$(\Lambda/\hbar)^{2|\lambda|}\bigl(\prod_{s \in \lambda}h(s)\bigr)^{-2}$, 
which is the standard weight of a random partition, 
that is,  
a Poissonized Plancherel measure of symmetric group. 
It can be also viewed as a melting crystal model, 
known as random plane partition.  
The corresponding model 
is studied in \cite{Nakatsu-Takasaki} 
as a melting crystal model with external potential, 
where the Chern characters $\mathcal{O}_k$ in (\ref{Z_U(1)}) 
correspond precisely to the external potentials.

\subsection{Melting crystal model with external potentials}

A plane partition $\pi$ is an array of 
non-negative integers 
\begin{eqnarray}
\begin{array}{cccc}
\pi_{11} & \pi_{12} & \pi_{13} & \cdots \\
\pi_{21} & \pi_{22} & \pi_{23} & \cdots \\
\pi_{31} & \pi_{32} & \pi_{33} & \cdots \\
\vdots & \vdots & \vdots & ~
\end{array}
\label{pi}
\end{eqnarray}
satisfying 
$\pi_{ij}\geq \pi_{i+1 j}$ and $\pi_{ij}\geq \pi_{i j+1}$ 
for all $i,j \geq 1$. 
Plane partitions are identified 
with the $3d$ Young diagrams. 
The $3d$ diagram $\pi$ 
is a set of unit cubes such that $\pi_{ij}$ cubes 
are stacked vertically on each $(i,j)$-element of $\pi$. 
%
%
\begin{figure}[ht]
\begin{center}
\includegraphics[scale=0.6]{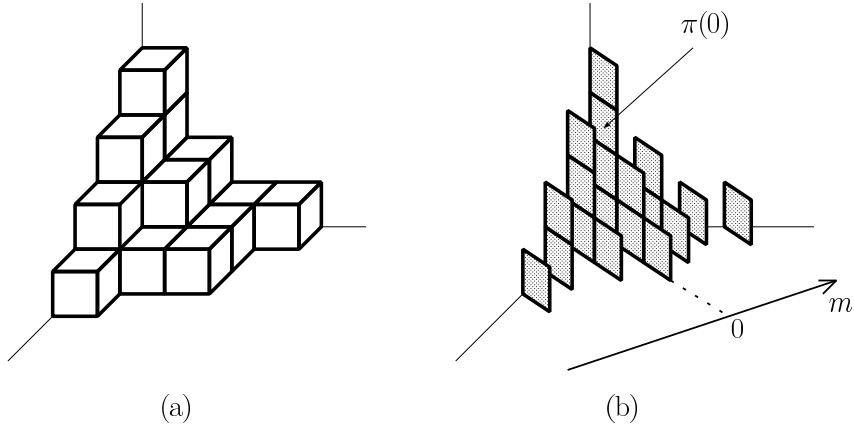}
\caption{\textit{
The $3d$ Young diagram (a) 
and the corresponding sequence of partitions 
 (b).}}
\end{center}
\label{three-dimensional Young diagram}
\end{figure}
%
%
Diagonal slices of $\pi$ become partitions, 
as depicted in Fig.1.  
Denote $\pi(m)$ the partition along the $m$-th diagonal slice, 
where $m \in \mathbb{Z}$. 
In particular, 
$\pi(0)=(\pi_{11},\pi_{22},\cdots)$ 
is the main diagonal one.  
This series of partitions satisfies the condition
\begin{eqnarray}
\cdots \prec \pi(-2) \prec \pi(-1) \prec 
\pi(0) \succ \pi(1) \succ \pi(2) \succ \cdots,
\label{interlace relations}
\end{eqnarray}
where 
$\mu \succ \nu$ means the interlace relation;  
$\mu \succ \nu$ 
$\Longleftrightarrow$ 
$\mu_1 \geq \nu_1 \geq \mu_2 \geq \nu_2 
\geq \mu_3 \geq \cdots$.

The Hamiltonian picture emerges 
from the above interlace relations, 
by viewing a plane partition as evolutions of partitions 
by the discrete time $m$. 
In particular, 
transfer matrix formulation 
using $2d$ complex free fermions 
is presented in \cite{Ok-Res},  
by taking advantage of 
the well-known realization of partitions 
in the fermion Fock space. 
Let 
$\psi(z)=\sum_{m \in \mathbb{Z}}\psi_mz^{-m-1}$ 
and 
$\psi^*(z)=\sum_{m \in \mathbb{Z}}\psi^*_mz^{-m}$ 
be complex fermions with the anti-commutation relations, 
$\left\{ \psi_m,\psi^*_n \right\}=\delta_{m+n,0}$ 
and 
$\left\{ \psi_m,\psi_n \right\}=\left\{ \psi_m^*,\psi_n^* \right\}=0$. 
Partitions can be realized 
as states of the fermion Fock space. 
For a partition $\lambda$,  
the corresponding state reads 
\begin{eqnarray}
|\lambda \rangle 
=\,
\pm 
\prod_{i=1}^{\infty} 
  \psi_{i-\lambda_i-1} 
  \psi^*_{-i+1}\,
|0\rangle\,, 
\label{partition_state}
\end{eqnarray}
where 
$|0\rangle$ denotes the Dirac sea 
with the vanishing $U(1)$ charge,  
and is defined by the conditions  
\begin{eqnarray}
\psi_m |0\rangle = 0
~~~
\mbox{for~ $\forall\, m \geq 0$}\,, 
~~~~~
\psi_m^* |0\rangle = 0 
~~~
\mbox{for~ $\forall\, m \geq 1$}\,. 
\end{eqnarray}

We may separate 
the interlace relations (\ref{interlace relations}) 
into two parts, 
each corresponding to 
(forward or backward) evolutions of partitions  
for $m \leq 0$ and $m \geq 0$ 
towards $\lambda=\pi(0)$.  
These two types of the evolutions are described 
in the transfer matrix formulation 
by using operators $G_{\pm}$ of the forms   
\begin{eqnarray}
G_{\pm}=
\exp
\Bigl\{
\sum_{k=1}^{\infty}
\frac{q^{\frac{k}{2}}}{k(1-q^k)}
J_{\pm k}
\Bigr\}, 
\label{G_pm}
\end{eqnarray}   
where 
$J_{\pm k}=
\sum_{n \in \mathbb{Z}}:\psi_{\pm k-n}\psi^*_n:$ 
are the modes of the $U(1)$ current 
$:\psi(z)\psi^*(z):
=\sum_{m \in \mathbb{Z}}J_mz^{-m-1}$. 
In accord with the 
types of the evolutions,  
$G_{\pm}$ generate partitions 
from the Dirac sea as 
\begin{eqnarray}
\langle 0|G_+=
\sum_{\lambda}s_{\lambda}(q^{\rho})
\langle \lambda|\,,
\hspace{8mm}
G_-|0 \rangle =
\sum_{\lambda}s_{\lambda}(q^{\rho})
|\lambda\rangle\,.
\label{G_pm_states}
\end{eqnarray}

In the free fermion description, 
we can convert 
the loop operators $\mathcal{O}_k$ 
to fermion bilinear operators $\hat{O}_k$ 
of the form  
\begin{eqnarray}
\hat{\mathcal{O}}_k=
(1-q^{-k})\sum_{n=-\infty}^{\infty}
q^{kn}:\psi_{-n}\psi^*_{n}: 
+1\,, 
\label{hat_O_k}
\end{eqnarray}
Actually, 
the state (\ref{partition_state}) 
becomes the simultaneous eigenstate and 
reproduces the Chern characters as the eigenvalues: 
\begin{eqnarray}
\hat{\mathcal{O}}_k|\lambda \rangle 
=
\mathcal{O}_k(\lambda)|\lambda \rangle\,. 
\label{eigenstate_O_k}
\end{eqnarray}
Therefore, 
taking account of (\ref{G_pm_states}) and (\ref{eigenstate_O_k}), 
we can express the generating function (\ref{Z_U(1)}) 
in the fermionic representation as 
\begin{eqnarray}
\mathcal{Z}^{U(1)}_{-i\hbar,i\hbar}(t)
=
\langle 0 |
G_+ 
Q^{L_0} 
\exp
\Bigl\{
\frac{1}{\hbar^2}\sum_{k=1}^{\infty}t_k \hat{\mathcal{O}}_k
\Bigr\} 
G_- 
|0\rangle\,, 
\label{fermionic representation}
\end{eqnarray}
where $L_0=\sum_{n \in \mathbb{Z}}n:\psi_{-n}\psi_n^*:$ is 
a special element of the Virasoro algebra.

\subsection{The integrable structure}

The fermion bilinears $\hat{\mathcal{O}}_k$ 
can be regarded as a commutative sub-algebra of 
the quantum torus Lie algebra 
realized by the free fermions 
\cite{Nakatsu-Takasaki}. 
The adjoint actions of $G_{\pm}$ 
on the Lie algebra generate automorphisms of the algebra. 
Among them, 
taking advantage of 
the shift symmetry, 
we can eventually convert  
the representation (\ref{fermionic representation}) 
into the expression \cite{Nakatsu-Takasaki} 
\begin{eqnarray}
\mathcal{Z}^{U(1)}_{-i\hbar,i\hbar}(t)
&=&
\langle 0 |
\exp
\Bigl\{
  \frac{1}{2\hbar^2}\sum_{k=1}^{\infty}(-)^k(1-q^{-k})t_kJ_k
\Bigr\}
\nonumber \\
&&
\hspace{12mm}
\times~
{\bf g}_{\star}^{5d\,U(1)}
\exp
\Bigl\{
\frac{1}{2\hbar^2}\sum_{k=1}^{\infty}(-)^k(1-q^{-k})t_kJ_{-k}
\Bigr\}
| 0 \rangle\,. 
\label{toda tau}
\end{eqnarray}
In the above formula, 
${\bf g}_{\star}^{5d\,U(1)}$ is the element of $GL(\infty)$ 
given by 
\begin{eqnarray}
{\bf g}_{\star}^{5d\,U(1)}=
q^{\frac{W}{2}}G_-G_+Q^{L_0}G_-G_+q^{\frac{W}{2}}\,, 
\label{g_U(1)}
\end{eqnarray}
where $W=W_0^{(3)}=\sum_{n \in \mathbb{Z}}n^2:\psi_{-n}\psi_n:$ 
is a special element of $W_{\infty}$ algebra. 
The loop operators $\mathcal{O}_k$ are converted 
to $J_k$ or $J_{-k}$ in the formula (\ref{toda tau}). 
Actually, 
$J_{\pm k}$ eventually become equivalent in the formula,  
since ${\bf g}_{\star}^{5d\,U(1)}$ 
satisfies the property \cite{Nakatsu-Takasaki}  
\begin{eqnarray}  
J_k\,{\bf g}_{\star}^{5d\,U(1)}={\bf g}_{\star}^{5d\,U(1)}J_{-k}\,, 
\hspace{6mm} 
\mbox{for}~ k \geq 0.
\label{reduction to 1-toda}
\end{eqnarray}

Viewing the coupling constants $t$ 
as a series of time variables, 
the right-hand side of (\ref{toda tau}) 
is the standard form 
of a tau function of $2$-Toda hierarchy 
\cite{Ueno-Takasaki}. 
However, 
by virtue of (\ref{reduction to 1-toda}), 
the two-sided time evolutions of $2$-Toda hierarchy 
degenerate to one-sided time evolutions. 
This is precisely the reduction to $1$-Toda hierarchy. 
Thus 
the generating function becomes a tau function of 
$1$-Toda hierarchy. 
The Toda hierarchy also has a discrete variable $s$, 
which corresponds to the $U(1)$ charge of the Dirac sea. 
The formula (\ref{toda tau}) is easily generalized 
to the cases, where the charge $s$ is interpreted as 
the vev of the Higgs field.

\section{Integral representation of energy functional}

A partition $\lambda=(\lambda_1,\lambda_2,\cdots)$ gives rise to 
a series of decreasing integers $\lambda_i-i$ $(i=1,2,\cdots)$. 
Vice versa, 
a series of decreasing integers $x_1>x_2>\cdots$, 
in which the $j$-th integer becomes 
$x_j=-j$ for $j > \exists i_0$, 
gives a partition. 
Such a series of decreasing integers 
is described by its density function as 
\begin{eqnarray}
\rho_{\lambda}(x)\,=\, 
\sum_{i=1}^{\infty} 
\delta(x-(\lambda_i-i))\,, 
\hspace{5mm} 
x \in \mathbb{R}\,. 
\label{rho_lambda(x)}
\end{eqnarray}
The statistical sum in the right-hand of (\ref{Z_U(1)}) 
can be converted to a statistical sum over the density functions. 
\begin{eqnarray}
\mathcal{Z}_{-i\hbar,i\hbar}^{U(1)}(t)\,=\, 
\sum_{\rho(\cdot)} 
e^{-\mathcal{E}[\rho(\cdot)]}\,, 
\label{energy function 1}
\end{eqnarray}
where $\mathcal{E}[\rho(\cdot)]$ is the energy functional 
obtained by taking the logarithm of the statistical weight.
\begin{eqnarray}
\mathcal{E}[\rho_{\lambda}(\cdot)]
\,=\,
-\log 
\left\{
Q^{|\lambda|}s_{\lambda}(q^{\rho})^2
e^{\hbar^{-2}\sum_{k=1}^{\infty}t_k\mathcal{O}_{k}(\lambda)}
\right\}\,. 
\label{energy functional 2}
\end{eqnarray}

The density function is rather implicit 
in the above expression of the energy functional. 
However, 
taking the consideration in \cite{Nekrasov-Okounkov}, 
we can convert the expression into 
an integration of a quadratic form of the density function. 
Prior to giving such an integral representation 
of the energy functional, 
let us describe the basic formula as follows: 
Let $f(x)$ be a function. 
For a partition $\lambda$, 
we put 
\begin{eqnarray}
F_{\lambda}\,=\, 
\sum_{s \in \lambda} f(h(s))
\,, 
\label{F_lambda}
\end{eqnarray}
where $h(s)$ denotes the hook length of the box $s \in \lambda$. 
The sum over the boxes in the right-hand side of (\ref{F_lambda}) 
can be converted to an integration of 
a quadratic form of the density function.  
To see this, 
we take a function $g(x)$ which satisfies 
\begin{eqnarray}
g(x+1)-2g(x)+g(x-1)&=&f(x)\,, 
\label{g(x)_condition_1}
\\
g(0)&=&0\,.  
\label{g(x)_condition_2}
\end{eqnarray}
By using such a function $g(x)$, 
we can eventually express $F_{\lambda}$ as 
\begin{eqnarray}
F_{\lambda}
\,=\, 
\frac{1}{2}
\int_{x \neq y}dxdy\,\,
g(|x-y|)\,
\Delta \rho_{\lambda}(x)
\Delta \rho_{\lambda}(y)\,, 
\label{formula:F_lambda}
\end{eqnarray}
where $\Delta$ 
denotes a difference operator of the form 
\begin{eqnarray}
\Delta \rho(x)\,\equiv \, 
\rho(x)-\rho(x-1)\,. 
\label{difference operator}
\end{eqnarray}
We provide a proof of the formula (\ref{formula:F_lambda}) 
in Appendix \ref{integral formula}.

As an application of the above formula, 
we describe an integral representation of 
the logarithm of the hook polynomial
\begin{eqnarray}
H_{\lambda}(q)\,=\, 
\prod_{s \in \lambda}(1-q^{h(s)})\,. 
\label{H_lambda_q}
\end{eqnarray}
Note that, 
by taking $f(x)=\log(1-q^x)$ in (\ref{F_lambda}), 
we find $F_{\lambda}=\log H_{\lambda}(q)$.
Therefore, 
by choosing a function $g(x\,;\,q)$ to satisfy   
the conditions 
(\ref{g(x)_condition_1}) and (\ref{g(x)_condition_2}),
where $f(x)=\log(1-q^x)$, 
the formula (\ref{formula:F_lambda}) 
gives the integral representation 
\begin{eqnarray}
\log H_{\lambda}(q)
&=& 
\frac{1}{2}
\int_{x \neq y}dxdy\,\,
g(|x-y|\,;\,q)\,
\Delta \rho_{\lambda}(x)
\Delta \rho_{\lambda}(y)\,.
\label{log_H_lambda(q)}
\end{eqnarray}
In the above formula, 
we can take $g(x\,;\,q)$  
to be the logarithm of the $q$-analogue 
of the Barnes $G$-function as 
\begin{eqnarray}
e^{g(x\,;\,q)}
\,=\,
(1-q)^{\frac{x(x-1)}{2}}
\mathbb{G}_2(x+1;q)\,. 
\label{g(x;q)}
\end{eqnarray}
To see that the above $g(x;q)$ 
actually satisfies 
(\ref{g(x)_condition_1}) and (\ref{g(x)_condition_2}), 
we sufficiently note that 
$\mathbb{G}_2(x;q)$ is the second cousin 
in the hierarchy of the multiple $q$-gamma functions 
$\mathbb{G}_n(x;q)$ ($n=0,1,\cdots$),  
which are defined by the following conditions \cite{Nishioka}: 
\begin{eqnarray}
\begin{array}{ll}
(i)
& 
{\displaystyle 
\mathbb{G}_n(x+1;q)=
\mathbb{G}_{n-1}(x;q)\mathbb{G}_n(x;q)\,,
}
\\[2.2mm]
(ii)
&
{\displaystyle
\mathbb{G}_n(1;q)=1\,,
} 
\\[2.2mm]
(iii)
&
{\displaystyle 
\frac{d^{n+1}}{dx^{n+1}}
\log \mathbb{G}_{n+1}(x+1;q) 
\geq 0\,, 
\hspace{6mm} x \geq 0\,,
} 
\\[2.2mm]
(iv)
&
{\displaystyle 
\mathbb{G}_0(x;q)=[x]_q\,,
}
\end{array} 
\end{eqnarray}
where $[x]_q=\frac{1-q^x}{1-q}$. 
The infinite product representation of 
$\mathbb{G}_2(x;q)$ follows from the above conditions as 
\begin{eqnarray}
\mathbb{G}_2(x+1;q)
\,=\,
(1-q)^{-\frac{x(x-1)}{2}}
\prod_{k=1}^{\infty}
\left\{
\Bigl(
\frac{1-q^{x+k}}{1-q^k}
\Bigr)^k 
(1-q^k)^x 
\right\}\,. 
\label{G_2(x;q) product formula}
\end{eqnarray}
The above infinite product representation 
leads to the expansion of $g(x;q)$ 
in positive powers of $q^x$ as 
\begin{eqnarray}
g(x;q)=
\sum_{n=1}^{\infty}
\frac{1}{n}
\left\{
\frac{x}{1-q^{-n}}
-
\frac{1}{(1-q^n)(1-q^{-n})}
\right\}
+
\sum_{n=1}^{\infty}
\frac{q^{nx}}{n(1-q^n)(1-q^{-n})}\,.
\label{g(x;q)_expansion}
\end{eqnarray}

Let us describe 
an integral representation of the energy functional 
(\ref{energy functional 2}). 
For this end, 
we first divide $\mathcal{E}$ into two parts 
\begin{eqnarray}
\mathcal{E}[\rho_{\lambda}(\cdot)]
&=&
\mathcal{E}_1[\rho_{\lambda}(\cdot)]+\mathcal{E}_2[\rho_{\lambda}(\cdot)]
\,,
\end{eqnarray}
where 
\begin{eqnarray}
\mathcal{E}_1[\rho_{\lambda}(\cdot)]
&=& 
-\log 
\Bigl\{ Q^{|\lambda|}s_{\lambda}(q^{\rho})^2\Bigr\}\,, 
\label{E_1_1}
\\ 
\mathcal{E}_2[\rho_{\lambda}(\cdot)]
&=&
-\hbar^{-2}\sum_{k=1}^{\infty}t_k
\mathcal{O}_{k}(\lambda)\,. 
\label{E_2_1}
\end{eqnarray}
To obtain an integral representation of $\mathcal{E}_1$, 
note that the $q$-hook formula (\ref{q-hook_formula}) 
enables us to factorize 
$Q^{|\lambda|}s_{\lambda}(q^{\rho})^2$ 
into the product 
$q^{-\kappa(\lambda)/2}\times 
\Bigl(\prod_{s \in \lambda}Qq^{h(s)}\Bigr)
H_{\lambda}(q)^{-2}$, 
where 
the logarithm of the first factor reads as 
\begin{eqnarray}
-\log q^{-\frac{\kappa(\lambda)}{2}}=
-\frac{\log q}{6} 
\int_{-\infty}^{+\infty}dx 
x^3 \Delta \rho_{\lambda}(x)\,,
\end{eqnarray}
while an integral representation of 
the logarithm of the second factor is easily 
obtained from the previous representation 
of $\log H_{\lambda}(q)$. 
By summing up these two, 
we eventually obtain the integral representation 
of $\mathcal{E}_1$ as 
\begin{eqnarray}
\mathcal{E}_1[\rho(\cdot)]
&=&
-\frac{\log q}{6} 
\int_{-\infty}^{+\infty}dx 
x^3 \Delta \rho(x) 
\nonumber \\
&&
+\int_{x \neq y}dxdy\,
g(|x-y| ; q,Q) \Delta \rho(x)\Delta \rho(y)\,, 
\label{E_1 2}
\end{eqnarray}
where $g(x;q,Q)$ is the logarithm of the following combination: 
\begin{eqnarray}
e^{g(x\,;\,q,Q)}\,=\, 
Q^{-\frac{x(x-1)}{4}}
q^{-\frac{x(x^2-1)}{12}}
(1-q)^{\frac{x(x-1)}{2}}
\mathbb{G}_2(x+1 ; q)\,. 
\label{g(x;q,Q)}
\end{eqnarray}
By using (\ref{G_2(x;q) product formula}), 
the expansion of $g(x;q,Q)$ 
in positive powers of $q^x$ takes the form
\begin{eqnarray}
g(x;q,Q)
&=&
\sum_{n=1}^{\infty}
\frac{1}{n}
\left\{
\frac{x}{1-q^{-n}}
-
\frac{1}{(1-q^n)(1-q^{-n})}
\right\}
-\frac{x(x-1)}{4}\log Q
-\frac{x(x^2-1)}{12}\log q 
\nonumber \\
&&
+
\sum_{n=1}^{\infty}
\frac{q^{nx}}{n(1-q^n)(1-q^{-n})}\,.
\label{g(x;q,Q)_expansion}
\end{eqnarray}

As regards the second part, we note 
\begin{eqnarray}
-\int_{-\infty}^{+\infty}dx 
q^{kx}\Delta \rho_{\lambda}(x)
&=& 
-\int_{-\infty}^{+\infty}dx 
q^{kx}
\Bigl(\rho_{\lambda}(x)-\rho_{\lambda}(x-1)\Bigr)
\nonumber \\
&=& 
\sum_{i=1}^{\infty}q^{k(\lambda_i-i+1)}
-\sum_{i=1}^{\infty}q^{k(\lambda_i-i)}\,. 
\end{eqnarray}
The right-hand side of this formula becomes a finite sum 
by cancellation of terms between the two sums and gives 
rise to $\mathcal{O}_k(\lambda)$. 
Thus we find 
\begin{eqnarray}
\mathcal{O}_k(\lambda)=
-\int_{-\infty}^{+\infty}dx 
q^{kx}\Delta \rho_{\lambda}(x)\,. 
\label{O_k(lambda)_integral}
\end{eqnarray}
Therefore, the integral representation of $\mathcal{E}_2$ 
becomes 
\begin{eqnarray}
\mathcal{E}_2[\rho(\cdot)]
&=&
\hbar^{-2}
\int_{-\infty}^{+\infty}dx\, 
\sum_{k=1}^{\infty}t_kq^{kx}
\Delta \rho(x)\,. 
\label{E_2 2}
\end{eqnarray}

\section{Thermodynamic limit and Riemann-Hilbert problem}


We consider the field theory limit of the $U(1)$ theory, 
which is achieved  by letting $\hbar \rightarrow 0$ 
and amounts to the thermodynamic limit of the melting crystal model 
or the $q$-deformed random partition.  
Actually, 
the statistical average of 
the number of boxes of a partition , 
where the partition is 
identified with the Young diagram 
in the standard manner,   
becomes of order $\hbar^{-2}$. 
This implies that, as $\hbar$ goes to zero, 
partitions that dominate are 
those of order $\hbar^{-2}$. 
To realize the thermodynamic limit, 
we have to rescale the Young diagrams 
by changing the size of each side of a box 
$\hbar$ times the original one. 
Correspondingly, 
the rescaling of partitions can be organized 
in terms of the density function, 
by rescaling $x$ to $u=\hbar x$ as 
\begin{eqnarray}
\rho(x=\frac{u}{\hbar})
\,=\,
\rho^{(0)}(u)+
O(\hbar)\,, 
\label{rho_(0)(u)}
\end{eqnarray}
where $\rho^{(0)}(u)$ denotes the scaled density function.

The statistical sum over the density functions 
in (\ref{energy function 1}) can be replaced, 
as $\hbar$ goes to zero, 
with the sum over the scaled density functions 
\begin{eqnarray}
\mathcal{Z}_{-i\hbar,i\hbar}^{U(1)}(t)
\,\simeq\,
\sum_{\rho^{(0)}(\cdot)} 
e^{-\frac{1}{\hbar^2}\mathcal{E}^{(0)}[\rho^{(0)}(\cdot)]}\,, 
\label{Z hbar->0 1}
\end{eqnarray}
where $\mathcal{E}^{(0)}$ denotes 
the classical energy functional 
obtained from $\mathcal{E}$ by  
\begin{eqnarray}
\mathcal{E}[\rho(x=\frac{u}{\hbar})]
\,=\, 
\frac{1}{\hbar^2}
\Bigl\{
\mathcal{E}^{(0)}[\rho^{(0)}(u)]+O(\hbar) 
\Bigr\}\,. 
\label{E_(0) 1}
\end{eqnarray}
Taking account of the expression (\ref{Z hbar->0 1}),  
the field theory limit is the semi-classical limit 
realized by minimizing 
the classical energy functional.

The classical energy functional
can be obtained from (\ref{E_1 2}) and (\ref{E_2 2}) 
by scaling the density function $\rho$ 
and the parameters as well.  
Noting that the difference operator $\Delta$ becomes 
the differential operator $\hbar d/du$ as $\hbar$ goes to zero,  
the classical energy functional takes the form 
\begin{eqnarray}
\mathcal{E}^{(0)}[\rho^{(0)}(\cdot)]
&=& 
\int_{u \neq v}dudv\, 
g^{(0)}(|u-v|\,;R,\Lambda) 
\frac{d\rho^{(0)}(u)}{du}
\frac{d\rho^{(0)}(v)}{dv}
+
\frac{R}{6}
\int_{-\infty}^{+\infty}du 
u^3
\frac{d\rho^{(0)}(u)}{du} 
\nonumber \\
&& 
+
\int_{-\infty}^{+\infty}du\,  
\sum_{k=1}^{\infty}
t_ke^{-kRu} 
\frac{d \rho^{(0)}(u)}{du}\,, 
\label{E_(0) 2}
\end{eqnarray} 
where $g^{(0)}(u;R,\Lambda)$ denotes 
the classical limit of $g(x;q,Q)$ as 
\begin{eqnarray}
g(x=\frac{u}{\hbar};q=e^{-R\hbar}, Q=(R\Lambda)^2)
\,=\,
\frac{1}{\hbar^2}
\Bigl\{ g^{(0)}(u;R,\Lambda)+O(\hbar) \Bigr\}\,. 
\label{g_(0)(u;R,Lambda) 1}
\end{eqnarray}
Explicitly, using the formula 
(\ref{g(x;q,Q)_expansion}), 
it is given by 
\begin{eqnarray}
g^{(0)}(u;R,\Lambda)
\,=\, 
\frac{R}{12}u^3
-\frac{\log R\Lambda}{2}u^2
-\frac{\zeta(2)}{R}u+\frac{\zeta(3)}{R^2}
-\frac{1}{R^2}\sum_{n=1}^{\infty}\frac{e^{-nRu}}{n^3}\,. 
\label{g_(0)(u;R,Lambda) 2}
\end{eqnarray}
The above function is subject to the following conditions: 
\begin{eqnarray}
&&
\frac{d^2g^{(0)}(u;R,\Lambda)}{du^2}
\,=\, 
\log 
\Bigl(
\frac{ \sinh \frac{Ru}{2}}{\frac{R\Lambda}{2}}\Bigr)\,,
\\[2mm]
&&
\frac{dg^{(0)}(0;R,\Lambda)}{du}
\,=\, 
g^{(0)}(0;R,\Lambda)
\,=\,0\,. 
\end{eqnarray}

Obtaining the thermodynamic limit reduces to 
solving the saddle point equation 
$\delta \mathcal{E}^{(0)}/\delta \rho^{(0)}(u)$ 
$=$ $0$,  
imposing the constraints \cite{Nekrasov-Okounkov} 
originally followed by $\Delta \rho_{\lambda}$. 
The variational problem can be  eventually 
summarized as the following integral equations: 
\begin{eqnarray}
\mbox{PP}
\int_{-\infty}^{+\infty}\!\!\!dv\,
2\partial_u
g^{(0)}(|u-v|\,;R,\Lambda) 
\frac{d\rho^{(0)}(v)}{dv}
&=&
-\frac{Ru^2}{2}-V'(u)
\hspace{7mm}
\mbox{on}~\mbox{supp}\bigl(\frac{d\rho^{(0)}}{du}\bigr)\,,
\label{variational problem 1}
\\[3mm]
\int_{-\infty}^{+\infty}\!\!\!du\, 
\frac{d\rho^{(0)}(u)}{du}
&=&-1\,, 
\label{variational problem 2}
\\[3mm]
\int_{-\infty}^{+\infty}\!\!\!du\, 
u\frac{d\rho^{(0)}(u)}{du}
&=&0\,, 
\label{variational problem 3}
\end{eqnarray}
where the integration symbol in the first equation 
means principle part, 
and we write the potential as 
\begin{eqnarray}
V(u)\,=\, 
\sum_{k=1}^{\infty}
t_ke^{-kRu}\,. 
\label{V(u)}
\end{eqnarray}

Solution of the variational problem 
gives the classical energy 
$\mathcal{E}_{\star}^{(0)}\equiv
\mathcal{E}^{(0)}[\rho^{(0)}_{\star}(\cdot)]$, 
where $\rho^{(0)}_{\star}$ denotes the solution.  
The vev of the loop operators $\mathcal{O}_k$ 
can be realized by 
the partial differentiations 
of $\mathcal{E}_{\star}^{(0)}(t)$ 
with respect to $t_k$. 
Actually, it is the vev 
at non-vanishing coupling constants $t$. 
We can express this quantity by using $\rho^{(0)}_{\star}$ 
in the form
\begin{eqnarray}
\frac{\partial \mathcal{E}^{(0)}_{\star}(t)}{\partial t_k}
\,=\,
\lim_{\hbar \rightarrow 0}
\langle \mathcal{O}_k \rangle 
\,=\, 
\int_{-\infty}^{+\infty}\!\!\!\!
du\, e^{-kRu}\frac{d\rho^{(0)}_{\star}(u)}{du}\,. 
\label{vev of O_k}
\end{eqnarray}
To see the above formula, 
note that eq.(\ref{variational problem 1}) 
can be integrated to 
\begin{eqnarray}
&&
\mbox{PP}
\int_{-\infty}^{+\infty}\!\!\!dv\,
g^{(0)}(|u-v|\,;R,\Lambda) 
\frac{d\rho^{(0)}_{\star}(v)}{dv}
\nonumber \\
&&
~~~~~~~~~~~~
=-\frac{1}{2}
\Bigl(\frac{Ru^3}{6}+V(u)\Bigr)
+\frac{1}{2}C(t)
\hspace{9mm}
\mbox{on}~\mbox{supp}\bigl(\frac{d\rho^{(0)}_{\star}}{du}\bigr)\,,
\label{variational problem 1_integrated}
\end{eqnarray}
where $C(t)$ is the integration constant.
By using the above equation and eq.(\ref{variational problem 2}), 
we can express $\mathcal{E}_{\star}^{(0)}$ as 
\begin{eqnarray}
\mathcal{E}_{\star}^{(0)}(t)
\,=\, 
-\int_{u \neq v}dudv\, 
g^{(0)}(|u-v|\,;R,\Lambda) 
\frac{d\rho^{(0)}_{\star}(u)}{du}
\frac{d\rho^{(0)}_{\star}(v)}{dv} 
-C(t)\,.
\label{E_(0)star}
\end{eqnarray}
The partial differentiation 
of the right-hand side of the above equation
with respect to $t_k$ 
leads to the formula (\ref{vev of O_k}).  
Actually, 
making use of eqs. 
(\ref{variational problem 1_integrated}) 
and (\ref{variational problem 2}), 
it can be computed as 
\begin{eqnarray}
\partial_{t_k}\mathcal{E}^{(0)}_{\star}(t)
&=& 
-2\int_{-\infty}^{+\infty}\!\!\!\!du\, 
\frac{d\rho^{(0)}_{\star}(u)}{du}
\partial_{t_k}
\left\{
\mbox{PP}\int_{-\infty}^{+\infty}\!\!\!\!dv\,
g^{(0)}(|u-v|;R,\Lambda)
\frac{d\rho^{(0)}_{\star}(v)}{dv}
\right\}
-\partial_{t_k}C(t)
\nonumber \\[1.5mm]
&=& 
-2\int_{-\infty}^{+\infty}\!\!\!\!du\, 
\frac{d\rho^{(0)}_{\star}(u)}{du}
\partial_{t_k}
\left\{
-\frac{1}{2}\Bigl(\frac{Ru^3}{6}+V(u)\Bigr)
+\frac{1}{2}C(t)
\right\}
-\partial_{t_k}C(t)
\nonumber \\[1.5mm]
&=& 
\int_{-\infty}^{+\infty}\!\!\!\!du\,
\partial_{t_k}V(u)
\frac{d\rho^{(0)}_{\star}(u)}{du}\,.
\end{eqnarray}
Thus, 
taking account of (\ref{V(u)}), 
we obtain the formula (\ref{vev of O_k}).

The variational problem can be reformulated  
as the Riemann-Hilbert problem to find out a certain analytic function.
To see this, 
consider the integral transform  
\begin{eqnarray}
\Phi(z)\,=\, 
\int_{-\infty}^{+\infty}du\, 
\coth 
\frac{R}{2}(z-u)\, 
\frac{d\rho^{(0)}(u)}{du}\, 
\hspace{8mm}
z \in \mathbb{C}\,. 
\label{Phi(z)}
\end{eqnarray}
The above function is actually an analytic function on 
the cylinder 
$\mathbb{C}^{*}=\mathbb{C}/\frac{2\pi i}{R}\mathbb{Z}$,  
since it is periodic with period $2\pi i/R$. 
The inverse-transform is realized by the imaginary part as 
\begin{eqnarray}
\frac{d\rho^{(0)}(u)}{du}
\,=\,
\mp \frac{R}{2\pi}
\Im \Phi(u\pm i0)\,. 
\label{inverse transform}
\end{eqnarray}

Therefore, 
taking account of the integral transformation (\ref{Phi(z)}), 
the Riemann-Hilbert problem is stated to obtain 
an analytic function on $\mathbb{C}^{*}-I$, 
where $I$ is an interval in the real axis, 
and satisfies the following conditions: 
\begin{eqnarray}
&&
\Re \Phi(u\pm i0)
\,=\,
-1-\frac{1}{R}V^{'''}(u) 
\hspace{8mm} 
\mbox{when}~u \in I\,, 
\label{RH 1}
\\[2mm]
&&
\Im \Phi(u\pm i0)
\,=\,0 
\hspace{8mm}
\mbox{when}~ 
u \in \mathbb{R}-I\,, 
\label{RH 2}
\\[2mm]
&&
\Phi(z)\rightarrow \mp 1
\hspace{8mm}
\mbox{as}~
\Re z \rightarrow \pm \infty\,, 
\label{RH 3}
\\[2mm]
&&
\oint_C dz\,z \Phi(z)=0\,
\hspace{5mm}(C:\mbox{a contour encircling $I$ anticlockwise})\,.
\label{RH 4}
\end{eqnarray} 
Among the above conditions, 
the first two correspond to the saddle point equation 
(\ref{variational problem 1}). More precisely, 
they are a twice differentiated form 
of the saddle point equation. 
The last two equations respectively amount to  
(\ref{variational problem 2}) and (\ref{variational problem 3}).

\section{Extended Seiberg-Witten geometry of $5d$ theory}

The solution of the forgoing Riemann-Hilbert problem 
was obtained in \cite{Maeda-Nakatsu},\cite{MNTT2} 
in the case where all the coupling constants vanish.  
By generalizing the argument there, 
we can solve the Riemann-Hilbert problem 
(\ref{RH 1})-(\ref{RH 4}), 
even in the case of non-vanishing coupling constants.

Let us employ the following curve 
which describes a fourth punctured $\mathbb{C}P_1$:
\begin{eqnarray}
\mathcal{C}_{\beta}~:~ 
y+y^{-1}\,=\, 
\frac{1}{R\Lambda}(e^{-Rz}-\beta)\,, 
\hspace{8mm} 
z \in \mathbb{C}\,, 
\label{C beta}
\end{eqnarray}
where $\beta$ is a real parameter satisfying the condition
$\beta > 2R\Lambda$. 
The above curve is an analogue of the so-called Seiberg-Witten 
hyperelliptic curve. Actually, 
it is a double covering of the cylinder 
$\mathbb{C}^{*}$, 
and $y$ is uniformized on the curve. 
See Fig. 2. 
In particular, 
the branch points of $y$ are located at 
$z=-\frac{1}{R}\log (\beta \pm 2R\Lambda)$. 
Thereby, 
$y$ has a single cut along the interval $I$ 
in the real axis, where 
$I=$
$\Bigl[
-\frac{1}{R}\log (\beta + 2R\Lambda)\,, 
-\frac{1}{R}\log (\beta -2R\Lambda)\,
\Bigl]$.
%
\begin{figure}[t]
\begin{center}
\includegraphics[scale=0.9]{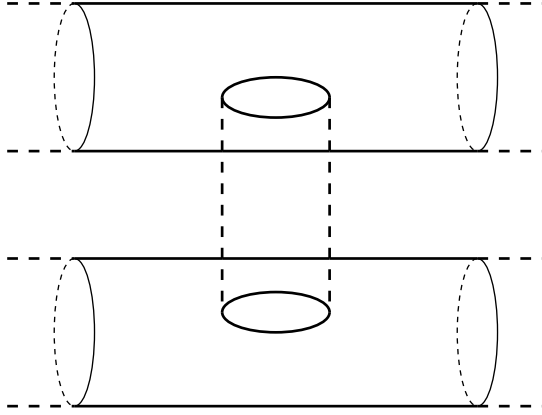}
\caption{\textit{
$\mathcal{C}_{\beta}$ is a double cover of cylinder.} }
\end{center}
\label{C_beta}
\end{figure}
%

We make an ansatz on the solution of the Riemann-Hilbert problem 
(\ref{RH 1})-(\ref{RH 4}) as  
\begin{eqnarray}
\Phi(z)\,=\,
-1+(2+\varphi(z)) 
\frac{1}{R}
\frac{d \log y}{dz}\,, 
\label{Phi(z) on C_beta 1}
\end{eqnarray}
where $\varphi(z)$ is a certain analytic function on $\mathcal{C}_{\beta}$. 
Through the above ansatz, 
conditions (\ref{RH 1})-(\ref{RH 4}) 
impose constraints on $\varphi(z)$. 
Actually, 
the first three conditions are translated as 
\begin{eqnarray}
&&
\Im \varphi(u\pm i0)
\,=\,
\left\{
\begin{array}{cc}
\displaystyle{
\pm \frac{1}{iR}V'''(u)
\sqrt{P(u-i0)} 
}
&
u \in I\,, \\[2mm]
0
&
u \in \mathbb{R}-I\,, 
\end{array}
\right. 
\label{varphi RH 1}
\\[4mm]
&&
\varphi(z) 
\,=\, O(1) 
\hspace{8mm} \mbox{as}~ \Re z \rightarrow +\infty\,, 
\label{varphi RH 2}
\\[2mm]
&&
\varphi(z)
\,=\, O(e^{Rz}) 
\hspace{8mm} \mbox{as}~ \Re z \rightarrow -\infty\,. 
\label{varphi RH 3}
\end{eqnarray}
In the above, 
$P(z)$ is a quadratic polynomial of $e^{Rz}$ given by 
\begin{eqnarray}
\frac{1}{R}d \log y 
\,=\, 
\frac{dz}{\sqrt{P(z)}}\,. 
\label{P(z)}
\end{eqnarray}
In other word, 
\begin{eqnarray}
P(z)
\,=\, 
(1-\beta e^{Rz})^2
-
(2R\Lambda e^{Rz})^2\,. 
\label{P(z)2}
\end{eqnarray}

An analytic function satisfying  
the above three conditions can be realized 
by a contour integral of the form  
\begin{eqnarray}
\varphi(z)
\,=\,
\frac{-1}{4\pi i}
\oint_C\!dw 
\Bigl(1+\coth \frac{R}{2}(z-w)\Bigr)
V'''(w)\sqrt{P(w)}\,, 
\label{varphi contour integral}
\end{eqnarray}
where $C$ is a closed curve that surrounds the interval $I$ 
but leaves $z$ outside as depicted in Fig. 3.
To see that the above satisfies 
the conditions (\ref{varphi RH 1})-(\ref{varphi RH 3}), 
note that the contour integration in the right-hand side 
of (\ref{varphi contour integral}) can be written as 
an integration along $I$ of the form 
\begin{eqnarray}
\varphi(z)
\,=\,
\frac{-1}{2\pi i}
\int_I\!dv 
\Bigl(1+\coth \frac{R}{2}(z-v)\Bigr)
V'''(v)\sqrt{P(v-i0)}\,.
\label{varphi_real_integral}
\end{eqnarray} 
When $z=u\pm i0$,
taking account of (\ref{P(z)2}), 
the imaginary part of the right-hand side of this equation 
becomes 
$\pm V'''(u)\sqrt{P(u-i0)}/iR $ if $u \in I$, 
otherwise vanishes. 
Therefore, 
$\varphi(z)$ given in (\ref{varphi contour integral}) 
obeys the condition (\ref{varphi RH 1}). 
The constant term of the kernel function in the contour integral  
(\ref{varphi contour integral}) 
is required in order to satisfy the boundary conditions 
(\ref{varphi RH 2}) and (\ref{varphi RH 3}). 
To see this, 
note that the kernel function has the asymptotics
\begin{eqnarray}
1+\coth \frac{R}{2}z 
&=& 
2\sum_{n=0}^{\infty}e^{-nRz}
\hspace{4mm}
\mbox{as}~\Re z \rightarrow \infty 
\nonumber \\
&=& 
-2\sum_{n=1}^{\infty}e^{nRz}
\hspace{4mm}
\mbox{as}~\Re z \rightarrow -\infty 
\,. 
\end{eqnarray}
By plugging the above 
into the right-hand side of (\ref{varphi_real_integral}), 
we find that $\varphi(z)$ satisfies 
the conditions (\ref{varphi RH 2}) and (\ref{varphi RH 3}).

%
%
\begin{figure}[t]
\begin{center}
\includegraphics[scale=0.85]{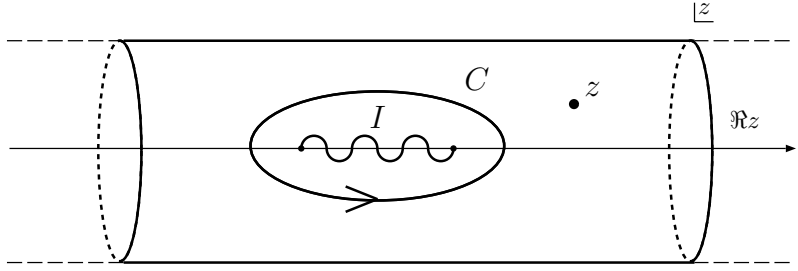}
\caption{\textit{
The cut $I$ and the contour $C$ 
on the Riemann sheet of $\mathcal{C}_{\beta}$.}}
\end{center}
\label{C_beta_sheet}
\end{figure}
%
%

\subsection{Determination of $\beta$}

In order to obtain the solution, 
taking $\varphi(z)$ as in (\ref{varphi contour integral}), 
we further need to impose the condition (\ref{RH 4}).  
This turns out to be an equation which 
determines the parameter $\beta$ of the curve (\ref{C beta}) 
in terms of $t$.

We can calculate the right-hand side of 
(\ref{varphi contour integral}) by residue calculus. 
To see this, we change the integration variable from $w$ 
to $W=e^{-Rw}$ as 
\begin{eqnarray}
\varphi(z)
\,=\,
\frac{1}{2\pi i}
\oint_{C'}\! 
\frac{dW}{W(W-e^{-Rz})}
\frac{1}{R}V'''(W)
\sqrt{(W-\beta)^2-(2R\Lambda)^2}\,, 
\label{varphi contour integral 2}
\end{eqnarray} 
where the contour $C'$, which is the image of $C$ 
by the mapping $w \mapsto W$, encircles 
the interval $[\beta-2R\Lambda,\, \beta+2R\Lambda]$ 
anticlockwise.  In the outside of $C'$, 
the integrand has poles at $W=e^{-Rz},\infty$ 
and is holomorphic elsewhere;
the origin $W=0$ is not a pole because 
$V'''(W)= \sum_{k=1}^\infty t_k(kR)^3W^k$ 
has a zero at $W=0$. 
We can deform $C'$ outward to small circles 
that surround the two poles clockwise.  Thus  
the contour integral reduces to a sum of 
the residues (times $-1$) at those poles, 
so that we have 
\begin{eqnarray}
\varphi(z)
\,=\,
-\frac{1}{R}V'''(z)\sqrt{P(z)}+M(z)\,. 
\label{varphi residue}
\end{eqnarray}
The first and second terms in the right-hand side 
are the contribution of the residues at 
$W=e^{-Rz}$ and $W=\infty$, respectively.  
More explicitly, 
\begin{eqnarray}
M(z)
&=&
\sum_{k=1}^{\infty}t_kM_k(z)\,, 
\label{M(z)}
\\
M_k(z)
&=&
-R^2k^3
\sum_{n=0}^kd_{k-n}e^{-nRz}\,, 
\hspace{8mm}
k=1,2,\cdots\,,
\label{M_k(z)}
\end{eqnarray}
where $d_n=d_n(\beta)$ denote the coefficients 
of expansion of $\sqrt{P(z)}$ in positive powers 
of $e^{Rz}$: 
\begin{eqnarray}
\sqrt{P(z)}
&=&
\sum_{n=0}^{\infty}
d_n(\beta)e^{nRz}
\hspace{8mm}
\mbox{as}~\Re z \rightarrow -\infty\,.
\label{d_n}
\end{eqnarray}
The first few terms read 
$d_0(\beta)=1\,, d_1(\beta)=-\beta\,, \cdots$.
Let us note that this expression of $M_k(z)$ 
and $M(z)$ readily implies that 
\begin{eqnarray}
R^2k^3e^{-kRz}\sqrt{P(z)}+M_k(z)
&=& O(e^{Rz}),\,
\\[2mm]
-\frac{1}{R}V'''(z)\sqrt{P(z)}+M(z)
&=& O(e^{Rz})  
\hspace{8mm}
\mbox{as}~\Re z \rightarrow -\infty\,.
\end{eqnarray}
We can thus directly confirm that 
the right-hand side of (\ref{varphi residue}) 
satisfies the boundary condition (\ref{varphi RH 3}).  
Actually, to solve the foregoing 
Riemann-Hilbert problem, we could have 
started from (\ref{varphi residue}) and 
proceeded to verifying the other conditions.

We now write down $\Phi(z)$, 
plugging the expression (\ref{varphi residue}) 
into (\ref{Phi(z) on C_beta 1}), as follows: 
\begin{eqnarray}
\Phi(z)
\,=\,
-\Bigl(1+\frac{1}{R}V'''(z)\Bigr)
+(2+M(z))\frac{1}{R}\frac{d \log y}{dz}\,. 
\label{Phi M}
\end{eqnarray}
By using this formula, 
the condition (\ref{RH 4}) becomes 
\begin{eqnarray}
\oint_C\!
z(2+M(z))d \log y\,=\,0\,. 
\label{RH 4_2}
\end{eqnarray}
This equation can be understood as an equation 
that determines the parameter $\beta$ of the curve 
(\ref{C beta}).
To write down such an equation explicitly, 
let us introduce the quantities 
\begin{eqnarray}
J_m(\beta)
&=& 
\oint_C\!
z e^{-mRz} d \log y\,, 
\hspace{8mm}
m=0,1,2,\cdots\,.
\label{J_m}
\end{eqnarray}
All the contour integrals that arise from 
components of $M(z)$ in $\oint_C\!z(2+M(z))d\log y$, 
taking account of (\ref{M(z)}) and (\ref{M_k(z)}),  
have the form (\ref{J_m}) 
and are consequently expressed 
in terms of $J_m(\beta)$ and $d_n(\beta)$. 
Eventually, 
they are arranged to provide 
the following equation for $\beta$:  
\begin{eqnarray}
\left(
2-R^2\sum_{k=1}^{\infty}k^3t_kd_k(\beta)
\right) 
J_0(\beta)
\,=\,
R^2\sum_{m=1}^{\infty}
\left(
\sum_{k=m}^{\infty}
k^3t_kd_{k-m}(\beta)
\right)
J_m(\beta)\,. 
\label{equation for beta}
\end{eqnarray}

The solution of the Riemann-Hilbert problem 
(\ref{RH 1})-(\ref{RH 4}) can be summarized as follows:
Take the curve $\mathcal{C}_{\beta=\beta(t)}$, 
where $\beta(t)$ is the solution of equation 
(\ref{equation for beta}). 
Then, 
the analytic function in (\ref{Phi M}) 
gives the solution.

\subsection{The case of single coupling constant}

\begin{figure}[t]
\begin{center}
\includegraphics[scale=1.08]{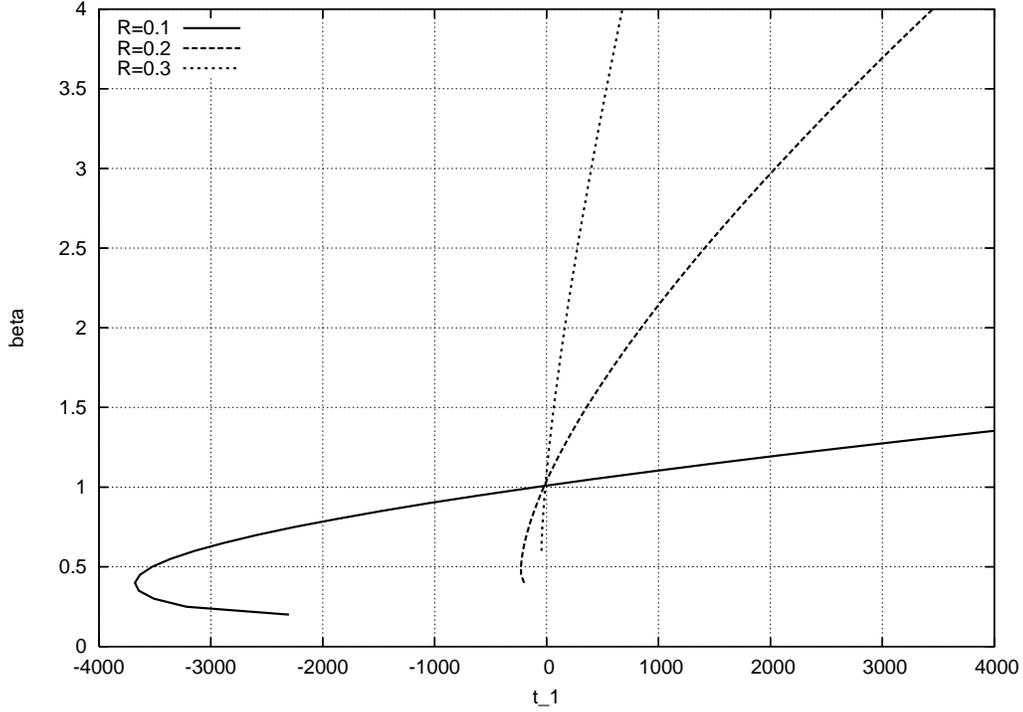}
\caption{\textit{
The graph of $\beta=\beta(t_1,0,0,\cdots)$.
Each corresponds respectively 
to the cases of $R=0.1$, $R=0.2$ and $R=0.3$, 
where $\Lambda$ is fixed to one.}}
\end{center}
\end{figure}

Let us examine the case of $t=(t_1,0,0,\cdots)$. 
In this case, 
eq.(\ref{equation for beta}) is simplified to be 
\begin{eqnarray}
(2+R^2t_1\beta)J_0(\beta)
\,=\,
R^2t_1J_1(\beta)\,, 
\label{equation for beta(t_1) 1}
\end{eqnarray}
where 
$J_{0,1}(\beta)$ 
are the integrals (\ref{J_m}). 
Contour integrations that appear in (\ref{J_m}) 
can be evaluated by computations 
utilizing the classical Jensen formula, 
and one eventually obtains 
\begin{eqnarray}
&&
J_0(\beta)\,=\,
\frac{2\pi i}{R}\log 
\Bigl(
\frac{\beta-\sqrt{\beta^2-(2R\Lambda)^2}}{2(R\Lambda)^2}
\Bigr)\,, 
\label{J_0(beta)}
\\[2mm]
&&
J_1(\beta)\,=\,
\frac{2\pi i}{R}
\left\{
  \beta \log 
       \Bigl(
         \frac{\beta-\sqrt{\beta^2-(2R\Lambda)^2}}{2(R\Lambda)^2}
       \Bigr)
  -\left(
         \beta-\sqrt{\beta^2-(2R\Lambda)^2}
   \right)
\right\}\,. 
\label{J_1(beta)}
\end{eqnarray}
We provide a derivation of the above formulas 
in Appendix \ref{Jensen's formula}. 
By plugging these values of the integrals 
into (\ref{equation for beta(t_1) 1}), 
eq.(\ref{equation for beta}) 
in the case of $t=(t_1,0,0,\cdots)$ becomes 
\begin{eqnarray}
t_1\,=\,
-\frac{2}{R^2}
\frac{1}{\beta-\sqrt{\beta^2-(2R\Lambda)^2}}
\log 
\left(
\frac{\beta-\sqrt{\beta^2-(2R\Lambda)^2}}{2(R\Lambda)^2}
\right)\,. 
\label{equation for beta(t_1) 2}
\end{eqnarray}

When the coupling constant $t_1$ vanishes, 
the above equation is further simplified to be 
$\beta-\sqrt{\beta^2-(2R\Lambda)^2}=2(R\Lambda)^2$. 
Taking account of the constraint $\beta > 2R\Lambda$, 
which is required to have a single cut along the real axis, 
one has a unique solution $\beta=1+(R\Lambda)^2$ 
only when $0<R\Lambda <1$. 
This is precisely 
the solution obtained 
in \cite{Maeda-Nakatsu},\cite{MNTT2}. 
Assuming that $0<R\Lambda <1$ as well,
one also has a unique solution 
$\beta(t_1)$ of eq.(\ref{equation for beta(t_1) 2})
such that $\beta(0)=1+(R\Lambda)^2$.
See Fig. 4.

\subsection{Seiberg-Witten differential}

The vev of the loop operator $\mathcal{O}_k$ 
can be represented by using an 
analogue of the so-called 
Seiberg-Witten differential.
Let $\Phi(z)$ be the foregoing solution 
described in the previous subsection. 
The inverse transform (\ref{inverse transform}) 
gives rise to the minimizer $\rho_{\star}^{(0)}$ as 
\begin{eqnarray}
\frac{d\rho_\star^{(0)}(u)}{du}
\,=\,
\left\{
\begin{array}{cc}
\displaystyle{
\mp\frac{1}{2\pi i}
(2+M(u))\frac{d \log y(u\pm i0)}{du}
}
&
u \in I\,,
\\[2mm]
0
&
u \in \mathbb{R}-I\,.
\end{array}
\right.
\label{rho_star M}
\end{eqnarray}
The integral expression 
(\ref{vev of O_k}) of the vev of $\mathcal{O}_k$ 
can be organized, 
by plugging the above formula into the expression 
and integrating by parts, 
to the following contour integral: 
\begin{eqnarray}
\frac{\partial \mathcal{E}^{(0)}_{\star}(t)}{\partial t_k}
\,=\,
\lim_{\hbar \rightarrow 0}
\langle \mathcal{O}_k \rangle 
\,=\, 
-\frac{1}{2\pi i}
\oint_C\!
kRe^{-kRz}
dS\,, 
\label{vev of O_k dS}
\end{eqnarray}
where 
$dS=S'(z)dz$ 
is an analogue of the Seiberg-Witten differential. 
$S'(z)$ is given by the indefinite integral 
\begin{eqnarray}
S'(z)\,=\,
-\int^z
(1+\frac{1}{2}M(z))d \log y
\,=\,
-\log y 
+N(z)\sqrt{P(z)}+\mathrm{const.}\,, 
\label{S'(z)}
\end{eqnarray} 
where $N(z)$ is a holomorphic function 
with an expansion similar to 
(\ref{M(z)}). 
Just like $\Phi(z)$, 
one can characterize $S'(z)$ 
by a Riemann-Hilbert problem, 
though we omit details.

The contour integral in the right-hand side of (\ref{vev of O_k dS}) 
can be converted to a residue integral. 
Actually, using coordinate $Z=e^{-Rz}$, we obtain 
\begin{eqnarray}
\frac{\partial \mathcal{E}^{(0)}_{\star}(t)}{\partial t_k}
\,=\,
\lim_{\hbar \rightarrow 0}
\langle \mathcal{O}_k \rangle 
\,=\, 
\mbox{Res}_{Z=\infty}
\Bigl(
kRZ^kdS
\Bigr)\,.
\label{vev of O_k dS 2}
\end{eqnarray}
The above expression shows that 
$\mathcal{E}_*^{(0)}(t)$ may be interpreted as a dispersionless 
tau function \cite{Krichever94}. 
This is natural because 
the partition function is substantially 
a tau function of 1-Toda hierarchy and 
$\mathcal{E}_*^{(0)}(t)$ gives 
the leading order part of the $\hbar$-expansion 
of $\log \mathcal{Z}_{-i\hbar,i\hbar}^{U(1)}(t)$.

\section{Conclusion and discussion}

We studied an extension of the Seiberg-Witten theory 
of $5d$ $\mathcal{N}=1$ SYM on $\mathbb{R}^4 \times S^1$.
We investigated correlation functions among 
the loop operators. 
These operators are analogues of the Wilson loop operators 
encircling the fifth-dimensional circle 
and give rise to the physical observables 
of topological-twisted  $5d$ $\mathcal{N}=1$ SYM 
in the $\Omega$ background 
through the equivariant descent equation. 
The correlation functions were computed 
by using the localization technique. 
Generating function of the correlation functions 
of $U(1)$ theory equals 
a statistical sum over partitions 
and reproduces the partition function of 
the melting crystal model with external potentials, 
where 
the loop operators are   
converted to the external potentials 
of the melting crystal model. 
This eventually shows that, 
by regarding the coupling constants of 
the loop operators as a series of time variables, 
the generating function is a $\tau$ function of $1$-Toda hierarchy. 
The $1$-Toda hierarchy therefore describes 
a common integrable structure of 
$5d$ $\mathcal{N}=1$ SYM in the $\Omega$ background 
and melting crystal model.

The thermodynamic limit of the partition function of this model 
was studied 
by applying an integral formula of the energy functional, 
and was reformulated 
as a Riemann-Hilbert problem 
to obtain an analytic function on 
$\mathbb{C}^*-I$, 
where 
$\mathbb{C}^*$ is a cylinder and 
$I$ is an interval in the real axis, 
and satisfies suitable conditions. 
We solved the Riemann-Hilbert problem 
that determines the limit shape of the main diagonal slice 
of random plane partitions in the presence of external 
potentials, and identified a relevant complex curve 
and the associated 
Seiberg-Witten differential, 
where the vev's of the loop operators 
are expressed as residue integrals 
of the differential.

Lastly, let us briefly discuss the issue of 
$4d$ limit (see also the discussion on this issue 
in our previous work \cite{Nakatsu-Takasaki}). 
It seems quite difficult to achieve a reasonable 
$4d$ ($R \to 0$) limit in the present framework.   
The difficulty is rather obvious in the setup 
of the Riemann-Hilbert problem.  As the definition 
(\ref{Phi(z)}) shows, it will not be $\Phi(z)$ 
but rather $R\Phi(z)$ that turns into 
the $4d$ counterpart 
\begin{eqnarray}
\Phi_{4d}(z) 
\,=\,
\int_{-\infty}^{+\infty}\, 
\frac{du}{z-u}\, 
\frac{d\rho^{(0)}(u)}{du}\ 
\end{eqnarray} 
of the integral transform of 
$\frac{d\rho^{(0)}(u)}{du}$ 
in the limit as $R \to 0$. 
If we multiply both hand sides of 
the first equation (\ref{RH 1}) 
of the Riemann-Hilbert problem 
by $R$ and let $R \to 0$, the outcome 
is the equation 
\begin{eqnarray}
\Re \Phi_{4d}(u\pm i0)
\,=\, 
\lim_{R\to 0}
\frac{R}{2}
\left(-1-\frac{1}{R}V^{'''}(u)\right) 
\,=\, 0 
\hspace{8mm} 
(u \in I)
\end{eqnarray}
that has no potential term.  Thus, 
at least in this naive prescription, 
the external potentials simply decouple 
in the $4d$ limit.  (As regards the constant term 
$-1$, it is natural that it disappears, because 
this term originates in a $5d$ Chern-Simons term 
\cite{MNTT1}.)  One might argue that this difficulty 
can be simply avoided by rescaling $t_k$'s as 
$t_k \to R^{-2}t_k$ before letting $R \to 0$.  
Actually, this does not resolve the problem, 
because the exponential functions $e^{-kRu}$ 
in the external potentials themselves 
become constant functions in this limit, 
so that the Riemann-Hilbert problem 
still takes a degenerate form: 
\begin{eqnarray}
\Re \Phi_{4d}(u\pm i0)
\,=\, 
\frac{1}{2}
\sum_{k=1}^\infty t_kk^3 
\hspace{8mm} 
(u \in I)\,.
\end{eqnarray}
Thus these naive procedures fail to generate 
the polynomial potentials $u^k$, $k = 1,2,3,\ldots$ 
\cite{Marshakov-Nekrasov} on the right-hand side 
of the Riemann-Hilbert problem. 
A possible remedy for this difficulty will be 
to extend the set of loop operators $\mathcal{O}_k$ 
to $k < 0$.  One can thereby derive polynomial 
potentials from a linear combination of 
exponential potentials as, say, 
\begin{eqnarray}
\left(\frac{e^{-Ru}-e^{Ru}}{-2R}\right)^k 
\;\to\; u^k 
\hspace{8mm}
(R \to 0)\,.
\end{eqnarray}
This, however, causes some other problems.  
First of all, we have to point out that 
almost all part of our foregoing consideration 
assumes, implicitly or explicitly, that $q$ is 
in the range $0 < q < 1$.  Flipping the signature 
of $k$ amounts to changing $q \to q^{-1}$.  
To introduce the loop operators $\mathcal{O}_k$ 
for both $k > 0$ and $k < 0$ simultaneously, 
one will be forced to consider the case where 
$q = 1$ (and $|q| = 1$ more generally), 
which can only be reached as a kind of 
singular limit letting $q \to 1$. 
This is also the case for identification of 
the integrable structure;  $q \to 1$ is 
a singular limit in which some properties of 
the quantum torus Lie algebra (crucial for 
identification of the integrable structure) 
break down \cite{Nakatsu-Takasaki}. 
Moreover, our method for solving 
the Riemann-Hilbert problem, too, 
has to be modified if the potential $V(z)$ 
contains both $e^{-kRz}$ and $e^{kRz}$.  
Thus the issue of the $4d$ limit is 
not straightforward in our present framework, 
raising a number of interesting questions.

\appendix
\subsubsection*{\underline{Acknowledgements}}
K.T is supported in part by Grant-in-Aid for Scientific Research 
No. 18340061 and No. 19540179.

\section{The $T^2$-action on $\mathcal{A}_E$}
\label{torus_action_on_A}

We treat integral curves of $V_{\epsilon_1,\epsilon_2}$ 
as curves on $\mathbb{R}^4$ in this appendix. 
Denote the integral curve passing through $x$ by $\gamma^x$.  
More precisely, it can be expressed 
as the solution of the differential equation  
\begin{eqnarray} 
\frac{dx(s)}{ds}=V_{\epsilon_1,\epsilon_2}(x(s))\,, 
\hspace{6mm}
x(0)=x\,. 
\label{gamma_x}
\end{eqnarray}

Let $A$ be a gauge potential on the $SU(N)$-bundle $E$ 
on $\mathbb{R}^4$. 
We consider the parallel transports 
along the integral curves. 
In particular, 
along the above $\gamma^x$, 
the parallel transport of the fibre $E_{x(s)}$ to $E_x$ 
is described by the holonomy operator 
$\mbox{P}e^{-\int_{x(s)}^{x}A}$, 
where the path-ordered integration is achieved  
from $x(s)$ to $x$ along $\gamma^x$. 
Such parallel transports provide endomorphisms 
$\tau_{\epsilon_1,\epsilon_2}(s)$ of $E$ 
by the formula 
\begin{eqnarray}
(\tau_{\epsilon_1,\epsilon_2}(s)\phi)(x)
=
\mbox{P}e^{-\int_{x(s)}^xA}\,
\phi(x(s))\,, 
\label{tau_epsilon}
\end{eqnarray}
where $\phi$ is a section of $E$. 
The above endomorphisms define a $T^2$-action on 
$\Omega^0(\mathbb{R}^4,E)$. 
In particular, they satisfy the relation  
$\tau_{\epsilon_1,\epsilon_2}(s)\cdot 
\tau_{\epsilon_1,\epsilon_2}(s')
=\tau_{\epsilon_1,\epsilon_2}(s+s')$. 
The infinitesimal action is therefore given by 
\begin{eqnarray}
{\bf t}_{\epsilon_1,\epsilon_2}
\cdot 
\phi 
=
\left.
\frac{d}{ds}
\tau_{\epsilon_1,\epsilon_2}(s)\phi\,
\right|_{s=0}\,, 
\label{t_epsilon_phi}
\end{eqnarray}
where ${\bf t}_{\epsilon_1,\epsilon_2}$ denotes 
the generator of $T^2$ that gives $V_{\epsilon_1,\epsilon_2}$. 
The right-hand side of (\ref{t_epsilon_phi}) can be computed 
by using (\ref{tau_epsilon}). 
To see this, 
note that $x(s)$ takes a form 
$x(s)=x+sV_{\epsilon_1,\epsilon_2}(x)+O(s^2)$ 
for a very small $s$, 
thereby we have 
$\mbox{P}e^{-\int_{x(s)}^xA}
=1+sV_{\epsilon_1,\epsilon_2}^{\mu}A_{\mu}(x)+O(s^2)$ 
and 
$\phi(x(s))=
\phi(x)+sV_{\epsilon_1,\epsilon_2}^{\mu}\partial_{\mu}\phi(x)
+O(s^2)$. 
By combining these two, 
we can write the formula (\ref{tau_epsilon}) as   
\begin{eqnarray}
(\tau_{\epsilon_1,\epsilon_2}(s)\phi)(x)
&=& 
\Bigl(
1+sV_{\epsilon_1,\epsilon_2}^{\mu}A_{\mu}(x)
+O(s^2)
\Bigr)
\Bigl(
\phi(x)+sV_{\epsilon_1,\epsilon_2}^{\mu}\partial_{\mu}\phi(x)
+O(s^2)
\Bigr)
\nonumber \\[1.5mm]
&=&
\phi(x)+
s\,
\iota_{V_{\epsilon_1,\epsilon_2}}
d_A\phi(x)+
O(s^2)\,,  
\end{eqnarray}
where $\iota_{V_{\epsilon_1,\epsilon_2}}$ means 
an operation of contraction with 
the vector field $V_{\epsilon_1,\epsilon_2}$.
Thus, 
by plugging the above 
into the right-hand side of (\ref{t_epsilon_phi}), 
we obtain 
\begin{eqnarray}
{\bf t}_{\epsilon_1,\epsilon_2}
\cdot 
\phi 
=
\iota_{V_{\epsilon_1,\epsilon_2}}
d_A\phi\,. 
\label{t_epsilon_phi_2}
\end{eqnarray}

Similarly, 
the endomorphisms induce a $T^2$-action on $\mathcal{A}_E$, 
which is the space of all the gauge potentials on $E$, 
by the formula
\begin{eqnarray}
d_{\tau_{\epsilon_1,\epsilon_2}(s)\cdot A}
\,\tau_{\epsilon_1,\epsilon_2}(s)
=
\tau_{\epsilon_1,\epsilon_2}(s)\,
d_A\,, 
\label{T_2_action_on_A}
\end{eqnarray}
where $d_A=d+A$ is a covariant differential on $E$. 
In the above formula, 
we use the same symbol to denote the $T^2$-action. 
By using the holonomy operator, 
the formula can be written in an explicit form as 
\begin{eqnarray}
d+ 
\left(\tau_{\epsilon_1,\epsilon_2}(s)\cdot A\right)(x)
=
\mbox{P}e^{-\int_{x(s)}^xA}
\bigl(
d+A(x(s))
\bigr)
\Bigl(
\mbox{P}e^{-\int_{x(s)}^xA}
\Bigr)^{-1}\,, 
\label{T_2_action_on_A_explicit}
\end{eqnarray}

Let us describe the infinitesimal form of the above action. 
For a very small $s$, 
the right-hand side of (\ref{T_2_action_on_A_explicit}) 
can be computed as 
\begin{eqnarray}
&&
\mbox{P}e^{-\int_{x(s)}^xA}
\bigl(
d+A(x(s))
\bigr)
\Bigl(
\mbox{P}e^{-\int_{x(s)}^xA}
\Bigr)^{-1}
\nonumber \\[1.8mm]
&&
=
\Bigl(
1+sV_{\epsilon_1,\epsilon_2}^{\mu}A_{\mu}(x)
+O(s^2)
\Bigr)
\Bigl\{
d+A(x)
+s
  \Bigl(V_{\epsilon_1,\epsilon_2}^{\nu}\partial_{\nu}A_{\mu}
        +A_{\nu}\partial_{\mu}V_{\epsilon_1,\epsilon_2}^{\nu}
  \Bigr)(x)dx^{\mu}
+O(s^2)
\Bigr\}
\nonumber \\[1.5mm]
&&
~~\times 
\Bigl(
1-sV_{\epsilon_1,\epsilon_2}^{\mu}A_{\mu}(x)
+O(s^2)
\Bigr)
\nonumber  \\[1.8mm]
&&
=d+A(x)
+s 
\Bigl(
V_{\epsilon_1,\epsilon_2}^{\nu}\partial_{\nu}A_{\mu}
-
V_{\epsilon_1,\epsilon_2}^{\nu}\partial_{\mu}A_{\nu}
+[V_{\epsilon_1,\epsilon_2}^{\nu}A_{\nu},\,A_{\mu}]
\Bigr)(x)dx^{\mu}
+O(s^2)
\nonumber \\[1.8mm]
&&
=d+A(x)
+s\,\iota_{V_{\epsilon_1,\epsilon_2}}F_A(x)
+O(s^2)\,, 
\label{infinitesimal_T_2_action_on_A}
\end{eqnarray}
where $F_A=dA+A\wedge A$ 
is the curvature two form. 
Thus, the formula (\ref{T_2_action_on_A_explicit}) 
reads as 
\begin{eqnarray}
\tau_{\epsilon_1,\epsilon_2}(s)\cdot A
=
A+s\iota_{V_{\epsilon_1,\epsilon_2}}F_A+O(s^2)\,. 
\end{eqnarray}
Therefore the infinitesimal form of the $T^2$-action 
becomes  
\begin{eqnarray}
{\bf t}_{\epsilon_1,\epsilon_2}
\cdot 
A 
&=&
\left.
\frac{d}{ds}
\tau_{\epsilon_1,\epsilon_2}(s)A\,
\right|_{s=0}
\nonumber  \\[1.5mm]
&=&
\iota_{V_{\epsilon_1,\epsilon_2}}
F_A\,. 
\label{t_epsilon_A}
\end{eqnarray}

\section{Proof of the identity (\ref{formula of O})}
\label{Proof:formula of O}

We first rewrite the $Q$-transformations 
(\ref{Q transform (A, psi)}) and (\ref{Q transform H}) 
in the forms 
\begin{eqnarray}
Q_{\epsilon_1,\epsilon_2}A(t)
&=&
\psi(t)\,, 
\nonumber \\
Q_{\epsilon_1,\epsilon_2}\psi(t)
&=&
d_{A(t)}\phi(t)
-\iota_{V_{\epsilon_1,\epsilon_2}}F_{A(t)}
-\frac{dA(t)}{dt}\,,
\nonumber \\
Q_{\epsilon_1,\epsilon_2}\phi(t)
&=&
-\iota_{V_{\epsilon_1,\epsilon_2}}\psi(t)\,.
\end{eqnarray}
By using the above expression, 
we can easily see that 
the combination 
$F_{A(t)}-\psi(t)+\phi(t)$ 
satisfies the identity 
\begin{eqnarray}
\bigl(
d_{A(t)}-\iota_{V_{\epsilon_1,\epsilon_2}}+Q_{\epsilon_1,\epsilon_2}
\bigr)
\Bigl(
F_{A(t)}-\psi(t)+\phi(t)
\Bigr)
=
\frac{dA(t)}{dt}\,. 
\label{equiv_Bianchi_identity}
\end{eqnarray}
The above identity may be interpreted as  
a loop space analogue of the equivariant Bianchi identity 
\cite{Berline_Getzler_Vergne}. 
Actually, 
when $A(t),\psi(t)$ and $\phi(t)$ are constant loops, 
that is, when they do not depend on $t$, 
(\ref{equiv_Bianchi_identity}) 
reduces to 
\begin{eqnarray}
\bigl(
d_{A}-\iota_{V_{\epsilon_1,\epsilon_2}}+Q_{\epsilon_1,\epsilon_2}
\bigr)
\Bigl(
F_{A}-\psi+\phi
\Bigr)
=0\,, 
\label{equiv_Bianchi_identity_2}
\end{eqnarray}
where the combination $F_{A}(x)-\psi(x)+\phi(x)$ 
is naturally identified with the $T^2$-equivariant 
curvature of the universal connection 
\cite{Losev-Marshakov-Nekrasov}, 
and (\ref{equiv_Bianchi_identity_2}) 
substantially describes the equivariant Bianchi identity.

Let us derive the formula (\ref{formula of O}) 
by using the identity (\ref{equiv_Bianchi_identity}). 
To do this, 
note that, 
since $\psi(t)$ is a Grassmann-odd one-form on $\mathbb{R}^4$, 
while $F_{A(t)}$ and $\phi(t)$ are Grassmann-even two- and zero-forms, 
we have 
\begin{eqnarray}
&&
\bigl(
d_{\epsilon_1,\epsilon_2}+Q_{\epsilon_1,\epsilon_2}
\bigr)
W(x;t_1,t_2)
\nonumber \\[1.5mm]
&&
\hspace{-3mm}
=
-\int_{t_2}^{t_1}ds\, 
  \tilde{W}(x;t_1,s)
  \wedge 
     \bigl(
      d_{\epsilon_1,\epsilon_2}+Q_{\epsilon_1,\epsilon_2}
     \bigr) 
     \Bigl(
      F_{A(s)}-\psi(s)+\phi(s)
     \Bigr)(x)
  \wedge W(x;s,t_2)
\,, 
\label{B_a}
\end{eqnarray}
where $\tilde{W}$ is another generalization of the 
path-ordered integral (\ref{W_(0)}), and is given by 
\begin{eqnarray}
\tilde{W}(x;\,t_1,t_2)
=\mbox{P}\exp
\Bigl\{-\int_{t_2}^{t_1}dt \big(F_{A(t)}+\psi(t)+\phi(t)\big)(x)
\Bigr\}\,.
\label{tilde_W}
\end{eqnarray}
Since $\mathcal{O}(x)$ is the trace of $W(x;R,0)$, 
using (\ref{B_a}), 
we can compute  
$\bigl(d_{\epsilon_1,\epsilon_2}
+Q_{\epsilon_1,\epsilon_2}\bigr)\mathcal{O}(x)$ as 
\begin{eqnarray}
&&
\bigl(
d_{\epsilon_1,\epsilon_2}+Q_{\epsilon_1,\epsilon_2}
\bigr)\mathcal{O}(x)
\nonumber \\
&&~~=
-\int_{0}^{R}dt\, \mbox{Tr}
  \Bigl\{ 
      \tilde{W}(x;R,t)
      \wedge 
        \bigl(
          d_{\epsilon_1,\epsilon_2}+Q_{\epsilon_1,\epsilon_2}
        \bigr)
        \Bigl(
          F_{A(t)}-\psi(t)+\phi(t)
        \Bigr)(x)
  \Bigr.
\nonumber \\
&&
\hspace{30mm}
  \Bigl.
      \wedge W(x;t,0)
  \Bigr\}\,.
\label{B_b}
\end{eqnarray}
By using (\ref{equiv_Bianchi_identity}), 
The right-hand side of (\ref{B_b}) can be further evaluated as 
\begin{eqnarray}
&&
\hspace{-5mm}
\bigl(
d_{\epsilon_1,\epsilon_2}+Q_{\epsilon_1,\epsilon_2}
\bigr)\mathcal{O}(x)
\nonumber \\[1.5mm]
&&
=
-\int_{0}^Rdt\, \mbox{Tr}
     \Bigl\{
       \tilde{W}(x;R,t)
       \wedge \frac{dA(x,t)}{dt}
       \wedge W(x;t,0)
     \Bigr\}
\label{B_c_1} \\[1.2mm]
&&
+\int_{0}^Rdt\, \mbox{Tr}
     \Bigl\{
        \tilde{W}(x;R,t)
        \wedge A(x,t)
        \wedge
          \Bigl(
            F_{A(t)}-\psi(t)+\phi(t)
          \Bigr)(x)
        \wedge W(x;t,0)
      \Bigr\}
\label{B_c_2} \\[1.2mm]
&&
-\int_{0}^Rdt\, \mbox{Tr}
      \Bigl\{
        \tilde{W}(x;R,t)
        \wedge 
           \Bigl(
             F_{A(t)}+\psi(t)+\phi(t)
           \Bigr)(x)
        \wedge A(x,t)
        \wedge W(x;t,0)
      \Bigr\}\,.
\label{B_c_3}
\end{eqnarray}
In the above, 
we can arrange the first term of the right-hand side as
\begin{eqnarray}
(\ref{B_c_1})
&=&
-\int_{0}^Rdt\, \mbox{Tr}
    \left\{
       \frac{dA(x,t)}{dt}
       \wedge W(x;t,0)
       \wedge W(x;R,t)
    \right\}
\nonumber \\[1.5mm]
&=&
-\int_{0}^Rdt \mbox{Tr}
    \left\{
       \frac{dA(x,t)}{dt}
       \wedge W(x;t,t-R)
\right\}\,. 
\nonumber 
\end{eqnarray}
Similarly, 
the second term and the third term can be arranged into 
\begin{eqnarray}
(\ref{B_c_2})
&=&
\int_{0}^Rdt\, \mbox{Tr}
     \left\{
        A(x,t)
        \wedge
          \Bigl(
            F_{A(t)}-\psi(t)+\phi(t)
          \Bigr)(x)
        \wedge W(x;t,t-R)
      \right\}\,,
\nonumber \\[1.5mm]
(\ref{B_c_3})
&=&-\int_{0}^Rdt\, \mbox{Tr}
      \Bigl\{
        A(x,t)
        \wedge W(x;t,t-R)
        \wedge 
           \Bigl(
             F_{A(t-R)}-\psi(t-R)+\phi(t-R)
           \Bigr)(x)
      \Bigr\}\,.
\nonumber 
\end{eqnarray} 
These two are particularly combined to give  
\begin{eqnarray}
(\ref{B_c_2})+(\ref{B_c_3})
=
-\int_{0}^Rdt\, \mbox{Tr}
      \left\{
        A(x,t)
        \wedge 
        \frac{dW(x;t,t-R)}{dt}
      \right\}\,.
\nonumber 
\end{eqnarray}

Therefore, 
by summing up (\ref{B_c_1}),(\ref{B_c_2}) and (\ref{B_c_3}),  
we find 
\begin{eqnarray}
&&
\bigl(
d_{\epsilon_1,\epsilon_2}+Q_{\epsilon_1,\epsilon_2}
\bigr)\mathcal{O}(x)
\nonumber \\[1.5mm]
&&
=
-\int_{0}^Rdt\, \mbox{Tr}
    \left\{
       \frac{dA(x,t)}{dt}
       \wedge W(x;t,t-R)
    \right\}
-\int_{0}^Rdt\, \mbox{Tr}
      \left\{
        A(x,t)
        \wedge 
        \frac{dW(x;t,t-R)}{dt}
      \right\}
\nonumber \\[1.5mm]
&&
=
-\int_{0}^Rdt\, 
\frac{d}{dt}
\left[
    \mbox{Tr}
    \Bigl\{
       A(x,t)\wedge W(x;t,t-R)
    \Bigr\}
\right]\,.
\end{eqnarray}
The last integral is integrated to be  
$\mbox{Tr}\,A(x,0)\wedge W(x;0,-R)-
\mbox{Tr}\,A(x,R)\wedge W(x;R,0)$,  
and becomes, by virtue of the periodicity, zero.  
Thus we obtain the formula (\ref{formula of O}).

\section{Proof of the formula (\ref{formula:F_lambda})}
\label{integral formula}

The formula (\ref{formula:F_lambda}) is stated as follows.

\begin{formula}
Let $f(x)$ be a function on $\mathbb{R}$. 
Choose a function $g(x)$ to satisfy the conditions 
(\ref{g(x)_condition_1}) and (\ref{g(x)_condition_2}). 
Then, for any partition $\lambda=(\lambda_1,\lambda_2,\cdots)$, 
the following identity holds: 
\begin{eqnarray}
\sum_{s \in \lambda}f(h(s))
=
\int_{x>y}dxdy\,
g(x-y)\, \Delta \rho_{\lambda}(x) \Delta \rho_{\lambda}(y)\,,  
\label{formula}
\end{eqnarray}
where $h(s)$ denotes the hook length of the box $s \in \lambda$, 
and $\Delta \rho_{\lambda}(x)=\rho_{\lambda}(x)-\rho_{\lambda}(x-1)$ 
is a difference of the density function (\ref{rho_lambda(x)}) 
of $\lambda$.
\end{formula}

To prove the above formula, 
note that 
the density functions of a partition $\lambda$ 
and its conjugate partition 
$\tilde{\lambda}=(\tilde{\lambda}_1,\tilde{\lambda}_2,\cdots)$, 
where the Young diagram $\tilde{\lambda}$
is obtained 
by flipping the Young diagram $\lambda$ 
over its main diagonal, 
satisfy the relation 
$\Delta \rho_{\tilde{\lambda}}(x)
=\Delta \rho_{\lambda}(-x)$. 
We can thereby rewrite the right-hand side of (\ref{formula}) as
\begin{eqnarray}
\int_{x>y}dxdy\,
g(x-y)\, \Delta \rho_{\lambda}(x) \Delta \rho_{\lambda}(y)
=
\int_{x>y}dxdy\,
g(x-y)\, \Delta \rho_{\lambda}(x) \Delta \rho_{\tilde{\lambda}}(-y)
\label{eq_2_proof}
\end{eqnarray}

We compute the right-hand side of (\ref{eq_2_proof}). 
For this end, 
it is convenient to introduce the truth function $\chi$ by 
\begin{eqnarray}
\chi(n)=
\left\{
\begin{array}{cl}
1 & \mbox{iff}~n \geq 1\,, \\
0 & \mbox{otherwise}\,.  
\end{array}
\right.
\label{truth_function_chi}
\end{eqnarray}
Owing to the $\delta$-functions in the density functions, 
the right-hand side of (\ref{eq_2_proof}) 
can be expressed by using the above $\chi$ as 
\begin{eqnarray} 
&&
\int_{x>y}dxdy\,
g(x-y)\, \Delta \rho_{\lambda}(x) \Delta \rho_{\lambda}(y)
\nonumber \\
&&
\hspace{-4mm}
=
\sum_{i,j=1}^{\infty}
g(\lambda_i+\tilde{\lambda}_j-i-j)
\chi(\lambda_i+\tilde{\lambda}_j-i-j) 
+
\sum_{i,j=1}^{\infty}
g(\lambda_i+\tilde{\lambda}_j-i-j+2)
\chi(\lambda_i+\tilde{\lambda}_j-i-j+2)
\nonumber \\
&&
-2
\sum_{i,j=1}^{\infty}
g(\lambda_i+\tilde{\lambda}_j-i-j+1)
\chi(\lambda_i+\tilde{\lambda}_j-i-j+1)\,.
\label{eq_3_proof}
\end{eqnarray}

We can further simplify (\ref{eq_3_proof}). 
To see this, 
note that 
the combination $\lambda_i+\tilde{\lambda}_j-i-j$ 
becomes $\leq -2$ when the pair $(i,j)$ 
is not a box of the Young diagram $\lambda$,  
while it becomes $\geq 0$ when the pair is a box of $\lambda$.  
More precisely, 
when the pair $(i,j)$ is a corner of $\lambda$, 
which means that $(i,j)$ is a box of $\lambda$ 
but neither $(i+1,j)$ nor $(i,j+1)$ is, 
the combination $\lambda_i+\tilde{\lambda}_j-i-j$ 
is equal to $0$, 
otherwise it takes $\geq 2$. 
Therefore, 
(\ref{eq_3_proof}) is eventually simplified as 
\begin{eqnarray}
&&
\int_{x>y}dxdy\,
g(x-y)\, \Delta \rho_{\lambda}(x) \Delta \rho_{\lambda}(y)
\nonumber \\
&&
=
\sum_{(i,j)\in \lambda}
\left\{
g(\lambda_i+\tilde{\lambda}_j-i-j+2)
-2g(\lambda_i+\tilde{\lambda}_j-i-j+1)
+g(\lambda_i+\tilde{\lambda}_j-i-j)
\right\}
\nonumber \\
&&
~+
\sum_{corners~of~\lambda}g(0)\,.
\label{eq_4_proof}
\end{eqnarray}
In the above, 
by virtue of the condition (\ref{g(x)_condition_1}), 
the first term of the right-hand side 
becomes $\sum_{(i,j)\in \lambda}f(h(i,j))$, 
while the second term vanishes by the condition 
(\ref{g(x)_condition_2}). Thus we obtain the formula 
(\ref{formula}).

\section{Derivation of eqs.(\ref{J_0(beta)}) and (\ref{J_1(beta)})}
\label{Jensen's formula}

We derive eqs. (\ref{J_0(beta)}) and (\ref{J_1(beta)}) 
by using the classical Jensen formula and its variant. 
Let $a$ be a complex number such that $|a| \neq 0,1$. 
The classical Jensen formula 
is an integration formula of the form \cite{Everest-Ward}
\begin{eqnarray}
\frac{1}{2\pi}
\int_0^{2\pi}d\theta 
\log |e^{i\theta}-a|
=
\log \max \bigl(1, |a| \bigr)\,.
\label{Jensen formula}
\end{eqnarray}
We also use the following variant of the above formula: 
\begin{eqnarray}
\frac{1}{2\pi}\int_0^{2\pi}d\theta 
(e^{ik\theta}+e^{-ik\theta})
\log |e^{i\theta}-a|
=
\left\{
\begin{array}{cl}
{\displaystyle -\frac{1}{k} \Re a^{-k}} & ~\mbox{iff}~|a|>1 \\[4.5mm]
{\displaystyle -\frac{1}{k} \Re a^k} &  ~\mbox{iff}~0<|a|<1 
\end{array}
\right.\,
\label{generalized Jensen formula}
\end{eqnarray}
where $k=1,2,\cdots$.

Let us first express $J_m(\beta)$ (\ref{J_m}) 
in terms of a contour integral on the $y$-plane. 
To do this, 
note that, 
taking account of the relation 
$e^{-Rz}=R\Lambda(y+y^{-1})+\beta$, 
the contour $C$ in the right-hand side of (\ref{J_m}) 
can be chosen so that it maps to the unit circle 
$|y|=1$ on the $y$-plane. 
We also note that, 
on the unit circle, 
by virtue of the condition $\beta > 2R\Lambda$, 
$R\Lambda(y+y^{-1})+\beta$ takes positive real numbers.
Thereby, the contour integral in the right-hand side of 
(\ref{J_m}) can be written as 
\begin{eqnarray}
J_m(\beta)=
\frac{-1}{R}
\oint_{|y|=1}
\frac{dy}{y}
\Bigl\{
R\Lambda(y+y^{-1})+\beta 
\Bigr\}^m
\log 
|R\Lambda(y+y^{-1})+\beta|\,.
\label{J_m(beta)_y}
\end{eqnarray}

The above integral 
can be converted to a combination of 
the phase integrals that appear in the formulas 
(\ref{Jensen formula}) and (\ref{generalized Jensen formula}). 
To see this, 
note that $R\Lambda(y+y^{-1})+\beta$ is factorized into
\begin{eqnarray}
R\Lambda(y+y^{-1})+\beta=
R\Lambda y^{-1}(y-\alpha)(y-\alpha^{-1})\,, 
\label{factorization}
\end{eqnarray}
where 
$\alpha 
\equiv 
-\beta/2R\Lambda-\sqrt{(\beta/2R\Lambda)^2-1}$. 
By plugging the above into the right-hand side of (\ref{J_m(beta)_y}), 
the integral becomes a sum of phase integrals of the form
\begin{eqnarray}
J_m(\beta)&=&
-\frac{i\log R\Lambda}{R}
\int_0^{2\pi}d\theta 
\Bigl\{R\Lambda(e^{i\theta}+e^{-i\theta})+\beta\Bigr\}^m
\nonumber \\
&&
-\frac{i}{R}
\int_0^{2\pi}d\theta 
\Bigl\{R\Lambda(e^{i\theta}+e^{-i\theta})+\beta\Bigr\}^m
\log|e^{i\theta}-\alpha|
\nonumber \\
&&
-\frac{i}{R}
\int_0^{2\pi}d\theta 
\Bigl\{R\Lambda(e^{i\theta}+e^{-i\theta})+\beta\Bigr\}^m
\log|e^{i\theta}-\alpha^{-1}|\,. 
\label{J_m(beta)_Jensen}
\end{eqnarray}
The above phase integrals can be computed by applying the formulas 
(\ref{Jensen formula}) and (\ref{generalized Jensen formula}). 
For the cases of $m=0,1$, 
we particularly obtain  
eqs.(\ref{J_0(beta)}) and (\ref{J_1(beta)}).


\end{document}